\documentclass[11pt,preprint]{aastex}

\usepackage{pdflscape}
\usepackage{graphicx}
\usepackage{color}
\usepackage{textcomp}
\usepackage{natbib}
\usepackage{tikz}

\slugcomment{Draft Version 2018 September 24}

\shorttitle{}
\shortauthors{}

\begin{document}


\title{A Systematic Study of Hale and Anti-Hale Sunspot Physical Parameters}

\author{Jing Li\altaffilmark{1}}
\affil{$^1$Department of Earth, Planetary and Space Sciences, University of California at Los Angeles, Los Angeles, CA  90095-1567}
\email{jli@igpp.ucla.edu}


\begin{abstract}
We present a systematic study of sunspot physical parameters  using full disk magnetograms from MDI/SoHO and HMI/SDO.  Our aim is to use uniform datasets and analysis procedures to characterize the sunspots, paying particular attention to the differences and similarities between ``Hale'' and ``anti-Hale'' spots.   Included are measurements of the magnetic tilt angles, areas,  fluxes and polarity pole separations for  4385 sunspot groups in Cycles 23 and 24 each measured, on average, at $\sim$66 epochs centered on meridian-crossing. The sunspots are classified as either ``Hale'' or ``anti-Hale'', depending on whether their polarities  align or anti-align with Hale's hemispheric polarity rule.  We find that (1) The ``anti-Hale'' sunspots constitute a fraction ($8.1\pm0.4$)\%  of all sunspots, and this fraction is the same in both hemispheres and cycles; (2) ``Hale'' sunspots obey Joy's law in both hemispheres and cycles but ``anti-Hale'' sunspots do not.  Three equivalent forms of Joy's law are derived; $\sin\gamma=(0.38\pm0.05)\sin\phi$; $\gamma=(0.39\pm0.06)\phi$; and $\gamma=(23.80\pm3.51)\sin\phi$, where $\gamma$ is the tilt angle and $\phi$ is the heliospheric latitude; (3) The average Hale sunspot tilt angle is $\overline{\gamma} = 5.49^\circ\pm0.09$;  (4) The tilt angles, magnetic fluxes and pole separations of sunspots are interrelated, with larger  fluxes correlated with larger pole separations and smaller tilt angles. We present empirical relations between these quantities. Cycle 24 is a much weaker cycle than Cycle 23 in sunspot numbers, cumulative magnetic flux, and average sunspot magnetic flux. The ``anti-Hale'' sunspots are also much weaker than ``Hale'' sunspots  in those parameters, but they share similar magnetic flux distributions and average latitudes. We characterize the two populations, and aim to shed light on the origin of ``anti-Hale'' sunspots.
\end{abstract}



\keywords{Sun: dynamo - Sun: general - Sun: interior - Sun: magnetic fields - Sun: sunspots}

\section{Introduction}
Sunspot groups are formed when magnetic flux tubes rise, likely from the tachocline between the convection and radiation zones \citep{1982A&A...113...99V}. Precisely how  they form and the details of their ascent are invisible to observers. However,  patterns of sunspot surface activity offer clues to the inner work of the global solar magnetic field. Sunspots are observed to consist of pairs with opposite magnetic polarities. They are generally elongated in the east-west direction and the leading polarities generally lean closer to the equator than the trailing polarities.  \citet{1919ApJ....49..153H} noted  that most leading spots  have opposite polarities in opposite hemispheres, and also that the sense of this hemispheric polarity pattern switches from cycle to cycle. This is known as Hale's hemispheric polarity rule, or simply ``Hale's law''. Separately, A. H. Joy noticed, and \citet{1919ApJ....49..153H} reported, that sunspot axes increasingly tilt with latitude, a trend known as Joy's law.  In addition to Hale's and Joy's laws, Sp\"{o}rer's law describes the steady decrease of sunspot latitude through the solar cycle, forming a ``butterfly diagram'' in a time vs.~latitude plot. Discovered by Richard Carrington around 1861, and refined by Gustav Sp\"{o}rer, the sunspot butterfly diagram is interpreted as the product of  the dynamo waves \citep{1955ApJ...122..293P} or the poloidal field stretching by  differential rotation \citep{1961ApJ...133..572B}.

Hale's, Joy's and Sp\"{o}rer's laws indicate that the solar magnetic fields change globally in a cyclic pattern. A widely-held description of  solar magnetic activity centers on the interplay between a global poloidal field  and differential rotation \citep{1955ApJ...122..293P,1961ApJ...133..572B,2000JApA...21..373C}. When the global field is dominantly a poloidal field, the Sun is in an activity minimum and few or no sunspots are visible.  As the global poloidal field is stretched into a toroidal field by differential rotation \citep{1998ApJ...505..390S},  the field lines are stretched in both latitudinal and radial directions, ultimately, giving rise to concentrated magnetic flux tubes. These flux tubes ascend due to magnetic buoyancy and, through the suppression of convection and of radiative output, appear as sunspots \citep{1955ApJ...121..491P}. At this stage, the surface of the Sun is occupied by more and more sunspots first at  high latitudes, gradually at mid- and low latitudes, as the solar activity enters a maximum phase. Near the equator, the leading polarities of sunspot pairs annihilate with their counterparts on the opposite hemisphere.  At the same time, the trailing polarities disintegrate and are transported to high latitudes by poleward meridional flows \citep{2010Sci...327.1350H}. This dissipation process eventually results in a reversal of the polar fields  at the height of  solar activity maximum. The global field gradually evolves back to an axisymmetric dipole in the second part of the solar cycle \citep{1998ApJ...501..866V,2014SSRv..186..491J} until the Sun enters the next activity minimum, with a  polar fields  opposite in sign to the previous minimum. As the new cycle starts, emerging sunspots have  polarities opposite from the previous cycle, so accounting for Hale's law.

The Coriolis force, acting on the rapid expanding flux ropes as they ascend through the convective zone,  is  probably responsible for the sunspot tilt \citep{1993A&A...272..621D,1994SoPh..149...23H}. If the magnetic flux is sufficiently strong in the overshooting region, the flux loop rises while having little interaction with the materials in the convection zone \citep{1982A&A...113...99V,2009LRSP....6....4F}. The sunspot tilt angle is determined by the Coriolis acceleration, $-2\omega(\phi)\sin\phi(\Delta s/\Delta t)$, where $\omega(\phi)$ is the sun's spin rate at latitude $\phi$ and $\Delta s/\Delta t$ is the average separation rate of magnetic footpoints. Assuming the acceleration is constant, the sunspot tilt angle can be approximated as 

\begin{equation}
\sin\gamma \sim \omega(\phi) \Delta t \sin\phi
\label{coriolis}
\end{equation}

\noindent where $\Delta t$ is the average time for  sunspot group emergence at the surface, and $\phi$ is the latitude \citep{1991ApJ...375..761W}. Equation (\ref{coriolis}) is observable given a large number of sunspots and has the form of Joy's law for an averaged sun's spin rate.   Indeed, considerable observational effort has been devoted to the determination of sunspot tilt angles and derive Joy's law \citep{1919ApJ....49..153H,1989SoPh..124...81W,1991SoPh..136..251H,2010A&A...518A...7D,2012Ge&Ae..52..999I,2012ApJ...745..129S,2012ApJ...758..115L,2013SoPh..287..215M}. 

The tilts of the sunspot magnetic axes provide the poloidal components needed for the global field to revert to an axisymmetric dipole configuration. For example, the contribution of an individual sunspot group to the solar axial dipole field may be expressed as $D_{ss}\propto s \Phi \sin\gamma \cos\phi $ where $s$ is the pole separation, $\Phi$ is the total flux, $\gamma$ is the tilt angle, and $\phi$ is the latitude \citep{2014SSRv..186..491J}. Surface flux transport simulations confirm the importance of the sunspot tilt angles in determining the polar field strength \citep{2010ApJ...719..264C,2015ApJ...808L..28J} but the disintegration and transport of magnetic fields  to the poles  are imperfectly understood \citep{2009SSRv..144...15P}. \citet{2010A&A...518A...7D} and \cite{2012Ge&Ae..52..999I} found that the mean normalized tilt angles are anti-correlated with the strength of the cycle.  A close connection between the tilt angle and the helicity was recently reported by \citet{2014SSRv..186..285P} and helicity has emerged as an important sunspot parameter to assess   coronal mass ejections \citep{2006ApJ...644..575Z}. 

Our main aim in the present work is to improve on  previous studies in several respects: 1) we utilize a uniform dataset which is obtained with MDI/SoHO and HMI/SDO. Both instruments is operated in a similar fashion and their performance is stable. 2) we employ systematic measurement procedures to determine the sunspot parameters. and 3)  we obtain averaged values of parameters from a large number of independent measurements as each spot rotates across the solar disk. Our dataset is thus relatively immune to systematic errors, and to fluctuations in parameters associated with temporal variability of the spots.   The paper is organized as follows: Section 2 describes the data and measurements, Section 3 shows the results, Section 4 gives a discussion while Section 5 is a  summary.

\section{Data and Measurements}
Our study is based on two main data sets.  First, we use full disk magnetograms from the  {\it Michelson Doppler Imager} (MDI) \citep{1995SoPh..162..129S} onboard the {\it Solar and Heliospheric Observatory} (SoHO) and the {\it Helioseismic and Magnetic Imager} (HMI) \citep{2012SoPh..275..207S} on the {\it Solar Dynamic Observatory} (SDO).  Second, we use sunspot records  from the daily ``USAF/NOAA Solar Region Summary'' composed by {\it Space Weather Prediction Center} (SWPC). The magnetic data span two decades from May 1996 to the present time covering two solar activity cycles (23 and 24), and one full magnetic cycle. The MDI data cadence is 96 minutes, while HMI's is 12 minutes. To maintain consistency between the magnetograms from the two instruments,  we use one HMI magnetogram every 96 minutes, and bin the data into 1024$\times$1024 pixels. Magnetograms from both instruments have the pixel size 2\arcsec$\times$2\arcsec. The magnetic strength obtained from MDI is reduced by a factor 1.4 for consistency with HMI data (c.f.~\cite{2012SoPh..279..295L}).

\begin{figure}[t]
\includegraphics[width=1.\textwidth]{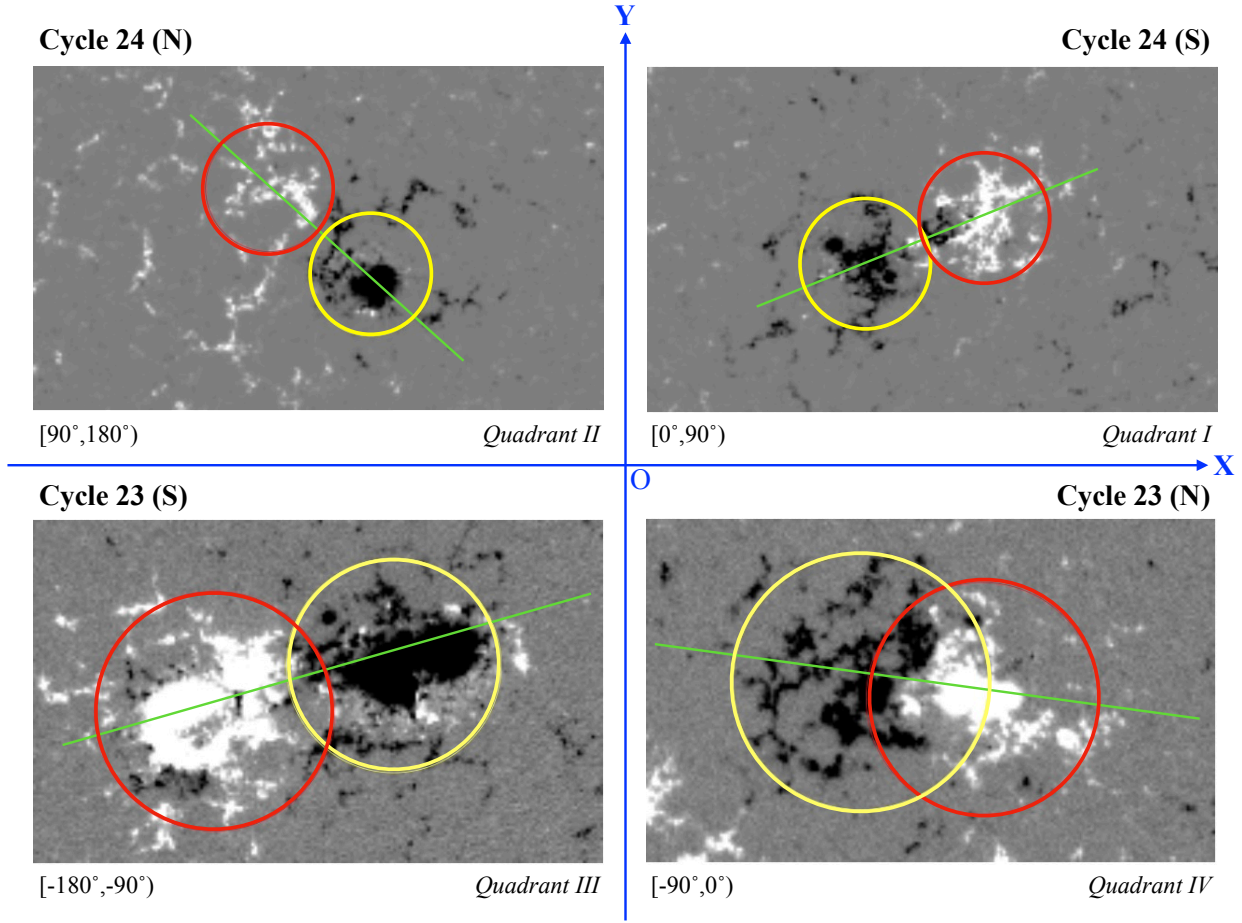}
\caption{Sunspot tilt angles are illustrated in Cartesian coordinates (axes are in blue colors). Four quadrants are filled with four actual sunspots which tilt angles agree with both Hale's and Joy's laws for given hemisphere and cycle. For example, Quadrant I is occupied by sunspots erupted in the southern hemisphere in Cycle 24 marked by Cycle 24 (S). The idealized bipolar magnetic pairs are represented by red and yellow circles,  over-plotted on actual sunspot magnetograms. Measured tilt angles are represented by green lines. The Roman numerals represent the quadrants. The tilt angle ranges are written in the lower left corner of each quadrant.  \label{tilt-definition}}
\end{figure}

We measured sunspot physical parameters  using automatic techniques similar to those employed in \citet{2012ApJ...758..115L}.  Our automatic program, written in IDL,  runs in the SolarSoftWare (SSW) environment \citep{1998SoPh..182..497F}. On each magnetogram, a sunspot is identified by its location as read from the USAF/NOAA sunspot records. The  area is outlined by an initial circle whose radius is 20$\times\sqrt{A/\pi}$ where $A$ is the sunspot ``total corrected area''. The initial circle is iteratively stretched into an ellipse and, finally, two circles having  opposite magnetic field polarities are positioned to fit the magnetogram data. Then the tilt angles, $\gamma$, are calculated using  

\begin{equation}
\tan\gamma=\Delta \phi /( \Delta \lambda \cos \bar \phi)
\end{equation}

\noindent where $\bar \phi$ is the mid-point latitude of the polarity pair, $\Delta \phi$ and $\Delta \lambda$ are the differences in latitude and longitude  between the centers of the two magnetic components.  Examples of the final fits are  illustrated in Figure (\ref{tilt-definition}) in which four sample sunspot magnetograms are shown. The best-fit sunspot polarity pairs are plotted as red and yellow circles with radii  calculated from $\sqrt{a/\pi}$ where ``a'' is best fit area of each component. Solid green lines in Figure (\ref{tilt-definition}) show the best-fit tilt axes, in each case. The four quadrants in the Figure distinguish sunspots whose tilt angles agree with both Hale's and Joy's laws in respective hemispheres and cycles.  

\begin{figure}[t]
\begin{center}
\includegraphics[width=0.72\textwidth]{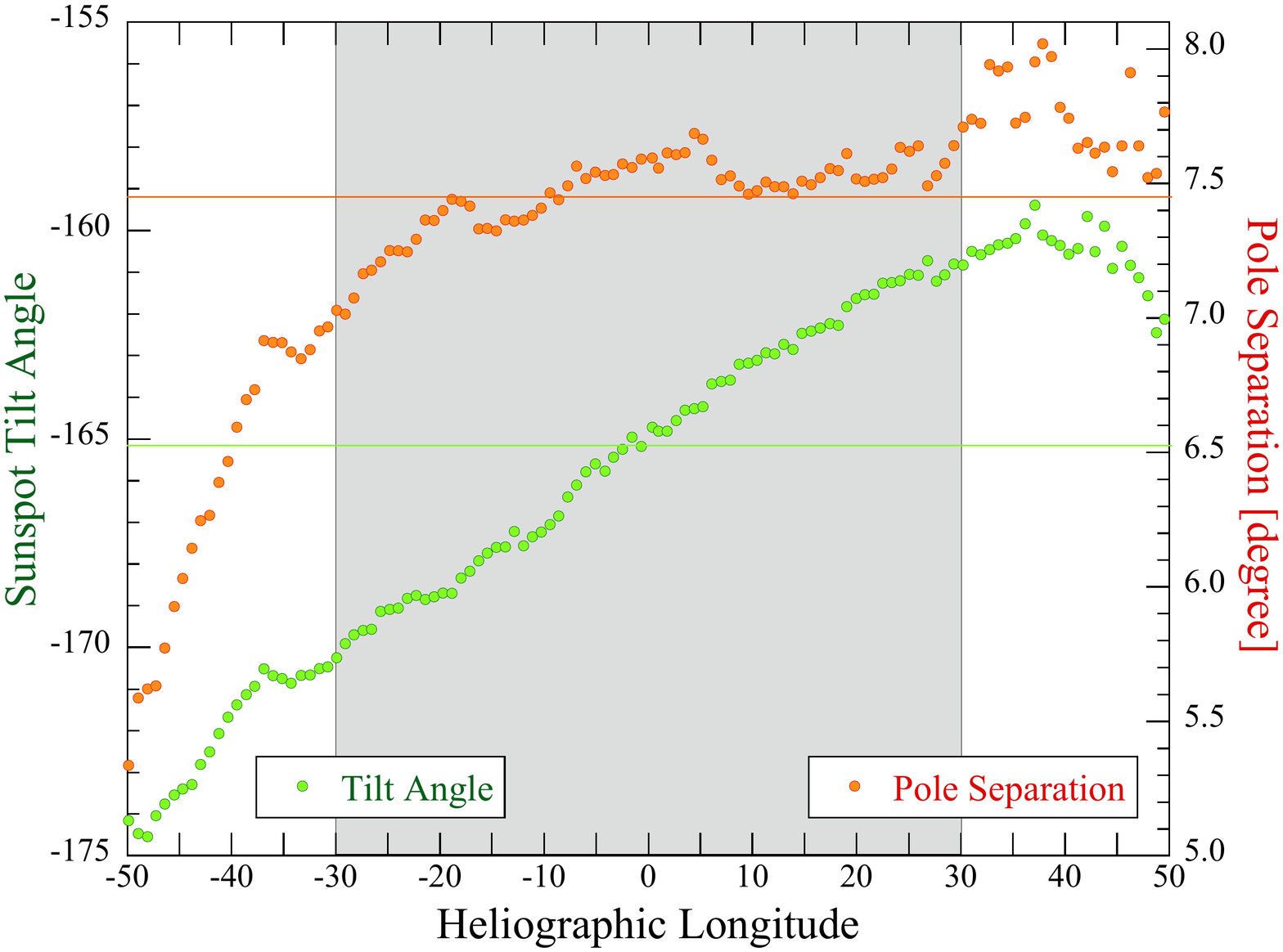}
\includegraphics[width=0.72\textwidth]{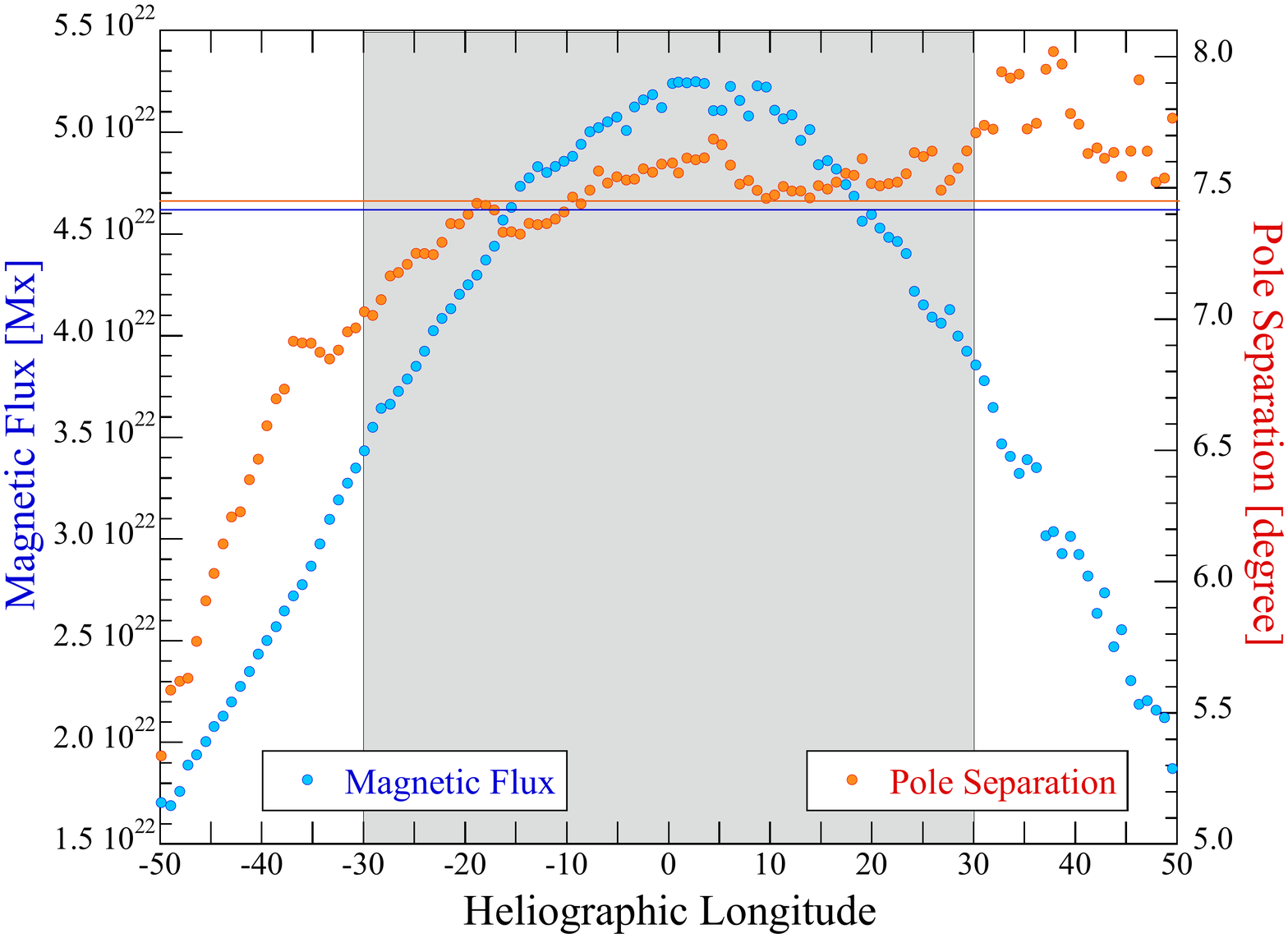}
\caption{An example of the sunspot parameters measured in a sunspot as a function of its heliographic longitude. The upper panel shows the tilt angle (green circles) and pole separation (red circles); the lower panel shows the total magnetic flux (blue circles) and pole separation (red circles) all as functions of the heliographic longitude. The grey shaded region indicates the longitudinal range used to calculate the average sunspot parameters. Red, green and blue horizontal lines represent the averaged pole separation, tilt angle and total magnetic flux of the sunspot group (NOAA10693). \label{example}}
\end{center}
\end{figure}

The automatic run does not guarantee a reliable set of sunspot tilt angles during the course of the disk passage. Two main effects lead to scatter in the tilt angles of a sunspot group.  First,  for very tiny sunspot groups, the code is unable to obtain convergent solutions for the north and south polarity components and so fails to produce reliable tilt angles.  Second, very close or overlapping sunspot groups cause the confusion with the procedure producing unreliable solutions.  We use the stability of repeated measurements to identify the problem sunspots.  Specifically, sunspots with tilt angles varying $>270^\circ$  in the course of their disk passages are targeted for inspection. The strategy greatly improved the sunspot tilt angle measurements along with other parameters.  About half of total sunspots went through this correction process. An example of the time dependence of parameters measured for an individual sunspot is shown in Fig. (\ref{example}).

\begin{figure}[t]
\includegraphics[width=1.0\textwidth]{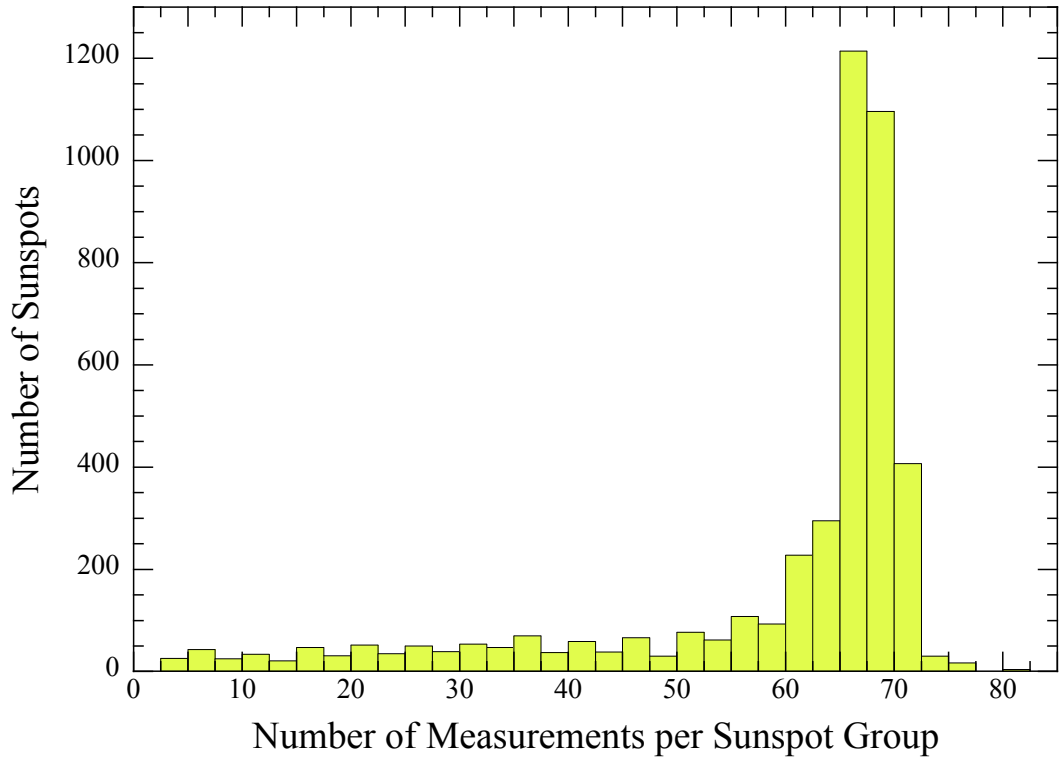}
\caption{A histogram of the sunspot counts as a function of the number of measurements per sunspot group. On average, each sunspot group has 66 measurements of the tilt angles, magnetic fluxes, pole separations and magnetic areas.  \label{histogram-ssn}}
\end{figure}

\begin{figure}[t]
\includegraphics[width=1.1\textwidth]{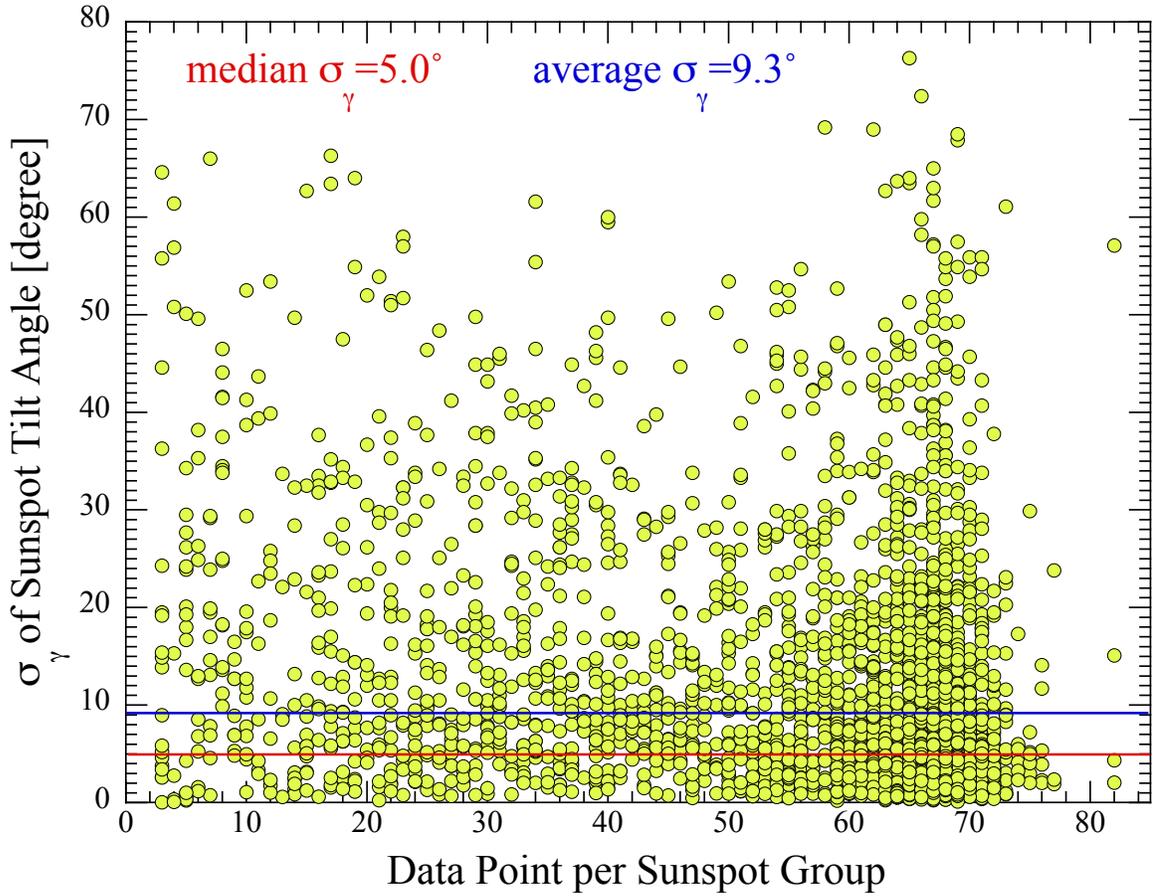}
\caption{Standard deviations of the  tilt angles for each sunspot, $\sigma_\gamma$, as a function of the number of available measurements. The red and blue horizontal lines represent the median and average standard deviation of tilt angles of all processed sunspots. \label{tilt-sigma}}
\end{figure}

To minimize projection effects, we consider only measurements of sunspots taken within $\pm30^\circ$ of meridian-crossing. This accounts for about 4.5-days of observation around the longitudinal disk center for each sunspot group (see the grey area in Fig. \ref{example}).  As shown in Figure (\ref{histogram-ssn}), a majority of sunspots are measured 60 to 70 times during this  period.  Figure (\ref{tilt-sigma}) shows the standard deviations, $\sigma_{\gamma}$, of the repeated tilt angle measurements as a function of the number of measurements for each sunspot group. The Figure shows two things.  First, the median and average $\sigma_{\gamma}$ are modest (5.0\degr~and 9.3\degr, respectively) indicating that the tilt angles are systematically variable.  Second, $\sigma_{\gamma}$ is independent of the number of measurements per spot, consistent with $\sigma_{\gamma}$ being a real measure of intrinsic temporal variations and not the result of statistical fluctuations in the measurements.   

Magnetograms are  two-dimensional maps of magnetic flux density. Our fitting procedure produces a pair of magnetic fluxes with opposite signs, for each sunspot on each  magnetogram. In the current work, each sunspot is represented by a set of parameters, including magnetic flux, averaged over the longitudinal range of $\pm30^\circ$.  The total sunspot magnetic flux means the unsigned magnetic flux, i.e., the sum of magnetic fluxes ($\Phi$) of opposite magnetic polarities. 

\begin{figure}[t]
\includegraphics[width=1.\textwidth]{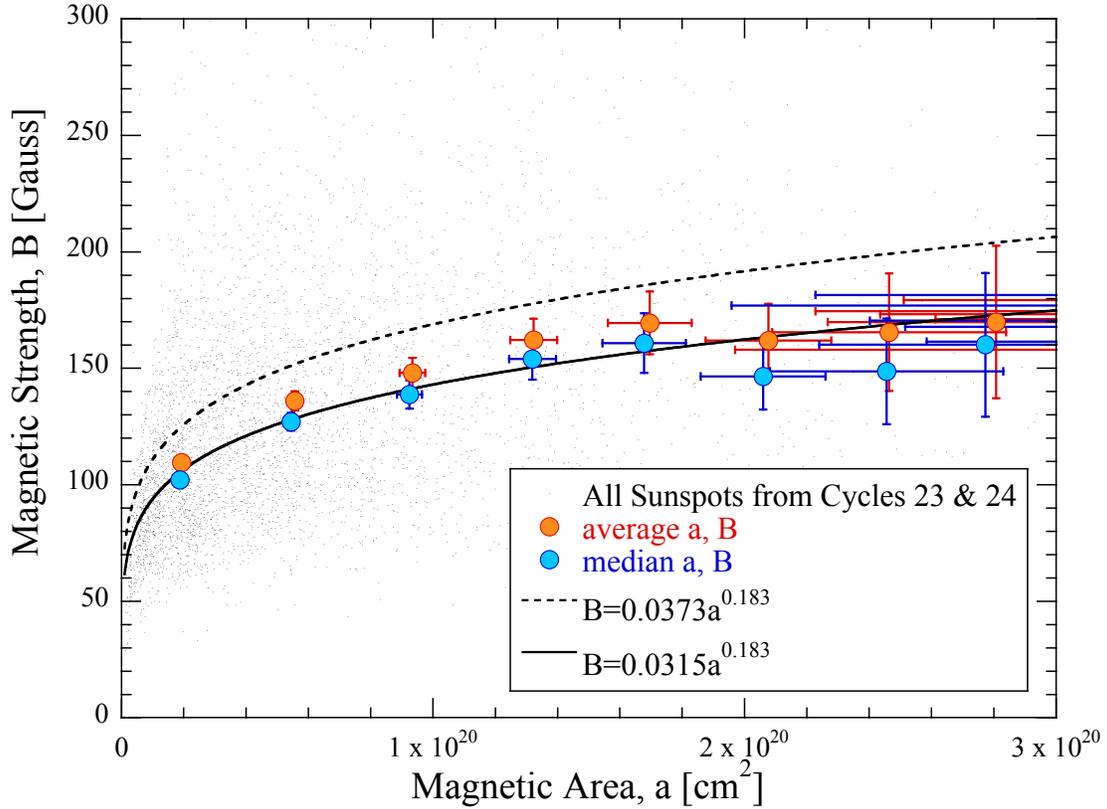}
\caption{The magnetic field strength as a function of sunspot magnetic area. Sunspots from both cycles are represented by symbols ``$\cdot$''. Filled orange and blue circles represent averaged and median magnetic area and strength in a $\sim3.8\times10^{19}$ [cm$^2$] bin. They are the guides to the eye, but are not used to obtain a fit. The solid curve is a direct power law fit to all sunspot data.  The dashed  curve, which is not a direct fit to the data,  is  obtained from Equation (\ref{eb2area}), itself differentiated from  Equation (\ref{eflux2area}). The difference in amplitudes (0.0373 vs.~0.0315) between the direct fitted equation and the analytical expression reflects the considerable uncertainty in $C_a$. \label{bfield}}
\end{figure}

The average magnetic strength of a sunspot group is defined by $B=\Phi/a$ [G], where ``$\Phi$'' is the unsigned magnetic flux [Mx];  ``$a$'' is the effective magnetic area [cm$^2$] computed from the total number of pixels in the idealized sunspot pairs.  The magnetic flux and area are highly linearly correlated (Pearson $r_{corr}=0.97$ at  P-Value $<10^{-5}$). We fitted a relation of the form to all sunspots

\begin{equation}
\log_{10}\Phi = k_a \log_{10} a+ C_a
\label{eflux2area}
\end{equation} 

\noindent to obtain $k_a=1.183\pm0.005$ and $C_a=-1.502\pm0.093$. When Equation (\ref{eflux2area}) is written as $\Phi(a)=\Phi_0 a^{k_a}$, where $\Phi_0=10^{C_a}$, we obtain the magnetic strength as a function of magnetic area:
\begin{equation}
B(a)=\frac{d\Phi(a)}{d a}=\Phi_0 k_a a^{k_a-1} 
\label{eb2area}
\end{equation}

\noindent which is shown in Figure (\ref{bfield}). All sunspots are plotted with ``$\cdot$''. The filled circles represent average (orange) and median (blue) values of the area ($a$) and magnetic strength ($B$) in $\sim3.8\times 10^{19}$ [cm$^2$] bin.   The average magnetic strengths are of order  10$^2$ Gauss, comparable to the magnetic field levels of plage regions visible in  Ca II emission \citep{1959ApJ...130..366L}. The dashed line is Equation (\ref{eb2area}) plotted with $k_a=1.183$ and $C_a=-1.502$; the direct non-linear least square fit to all sunspots is plotted with the solid curve. There is a  factor of 1.2 in amplitude between two curves (see equations in the figure). This is due to the large uncertainty, 0.093, in $C_a$ obtained from fitting Equation (\ref{eflux2area}). The uncertainty causes the amplitude for the analytical equation (\ref{eb2area}) varying from 0.030 and 0.046.

\section{Results}
\begin{figure}[t]
\includegraphics[width=1.1\textwidth]{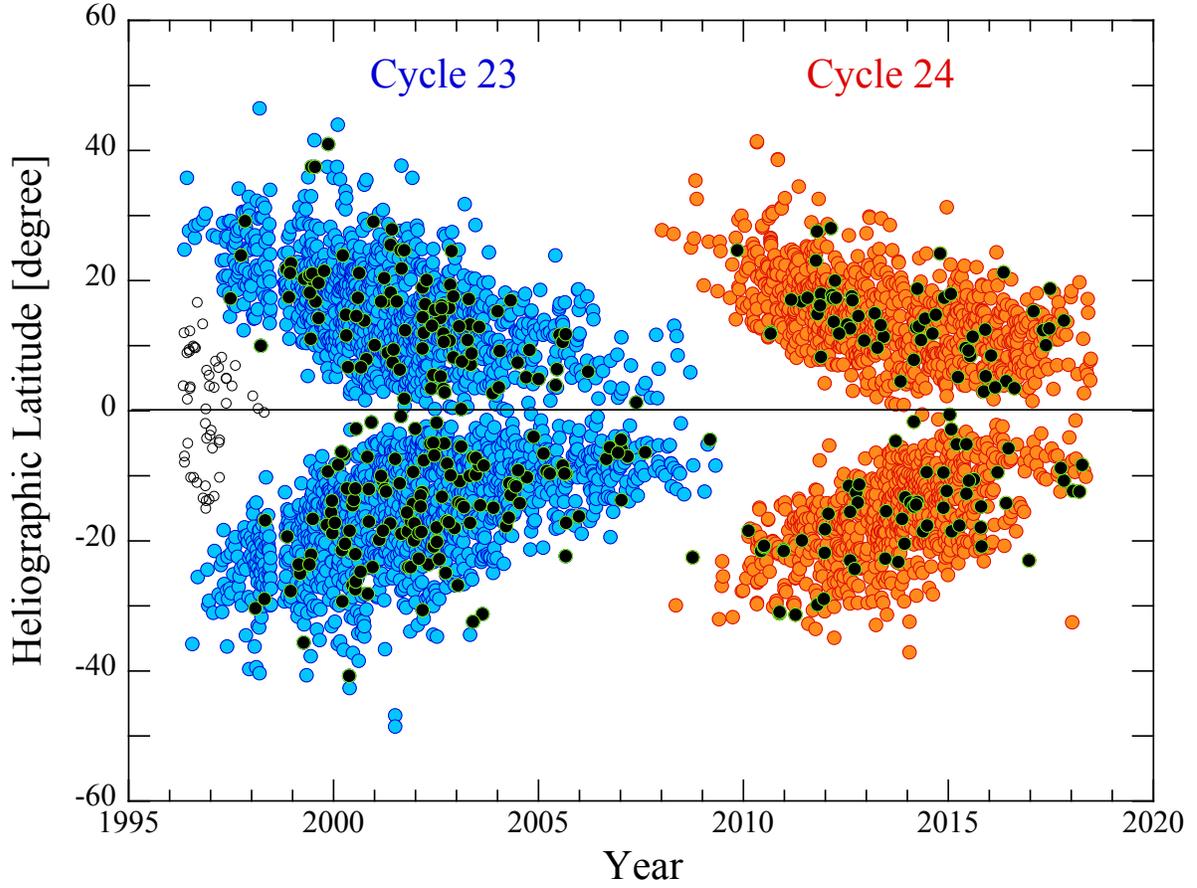}
\caption{Sunspot butterfly diagram (Sp\"{o}rer's law) for Cycles 23 and 24. Each circle represents a sunspot group erupted between May 1996 and July 2018: color blue is for Cycle 23 and red is for Cycle 24. The black filled dots with green circles represent the ``anti-Hale'' sunspots. Horizontal line marks the latitude $0^\circ$. Sunspots erupted at the end of Cycle 22 are marked with ``$\circ$'';  they are not examined in the current paper. \label{butterfly}}
\end{figure}

Figure (\ref{butterfly}) shows the Butterfly diagram computed from our data  in the time period  May 1996 to July 2018. Sunspots erupted in Cycles 23 and 24 are plotted with blue and orange filled circles, respectively.   The ``anti-Hale'' sunspots are plotted with black ``$\bullet$'' circled in green.  The diagram is consistent with the accepted onset of Cycle 24 starting at the end of 2008 \citep{2017SSRv..210..351W}. Sunspots erupted in Cycle 22 are plotted with  ``$\circ$'' symbols and excluded from further study here.


\subsection{Sunspot Hemispheric Asymmetry}
Table (\ref{thaleslaw}) lists the numbers of sunspots with respective hemispheres and cycles. It shows that the numbers of sunspots are distributed asymmetrically between the hemispheres. In Cycle 23, the number of sunspots in the southern hemisphere, $N_s(23)$, is greater than the number in the northern hemisphere, $N_n(23)$, but this asymmetry is reversed in Cycle 24. 
Specifically, the ratios are  $N_s(23)/N_n(23) = 1.20\pm0.03$ for Cycle 23 and $N_n(24)/N_s(24) =1.15\pm0.04$ for Cycle 24, where the listed uncertainty assumes Poisson statistics, $ratio/\sqrt{N_{min}}$ with $N_{min}$ the smaller of two sunspot numbers.  The average hemispheric asymmetry \textit{within one cycle} is $1.18\pm0.03$. On the other hand, the ratio $\frac{N_s(23)N_s(24)}{N_n(24) N_n(23)}  = 1.04\pm0.04$, is consistent with unity, meaning that the sunspot counts are hemisphere-symmetric over the (22 year) magnetic cycle of the Sun, to within the uncertainties of measurement.  

More sunspots erupted in Cycle 23 than in Cycle 24, with the ratio of the total numbers being $N(23)/N(24) = 1.56\pm0.04$ where, again, we quote a Poisson error bar. 

\subsection{Hale's Law}
\begin{figure}
\includegraphics[width=1.0\textwidth]{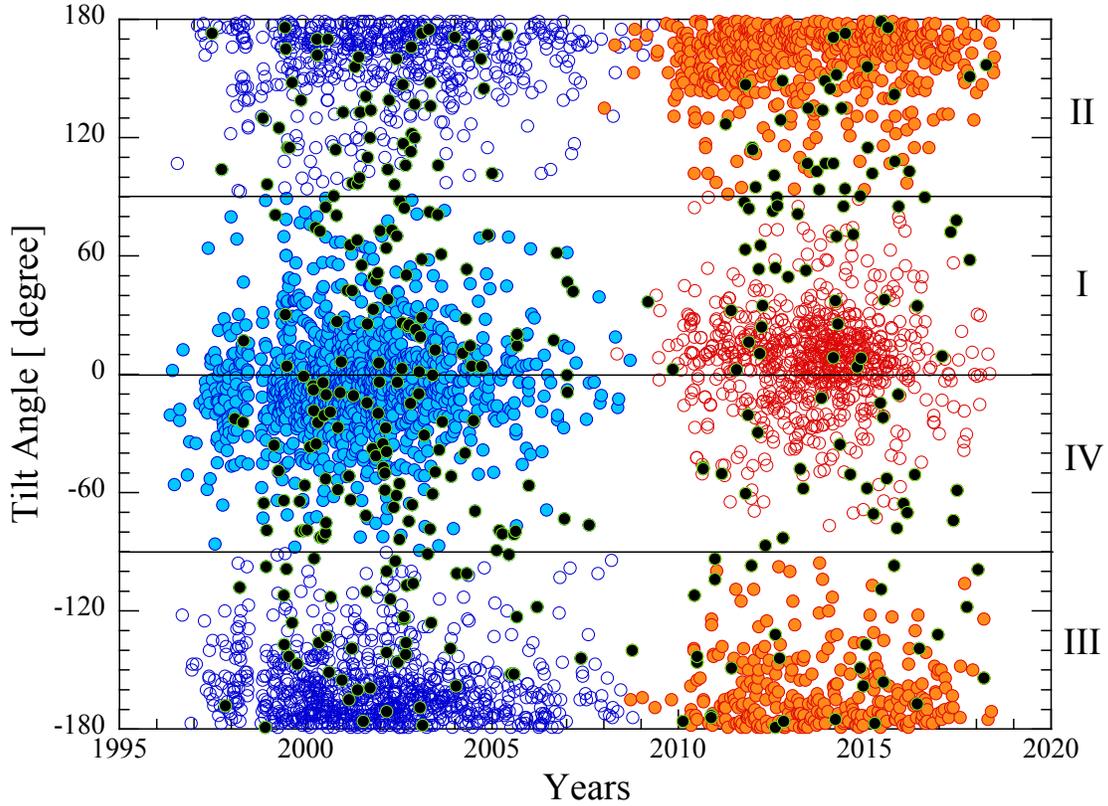}
\caption{Illustration of Hale's law. The sunspot tilt angles are plotted against time (year). Filled circles represent sunspots in the northern hemisphere, and empty circles represent those in the southern hemisphere. The ``$\bullet$'' with green circles represent the ``anti-Hale'' sunspots. Blue and red colors represent sunspots from Cycles 23 and 24, respectively.  Horizontal lines mark tilt angles $0^\circ$, $90^\circ$, and $-90^\circ$. The Roman letters on the right mark the quadrant numbers in Cartesian coordinates (see Fig. \ref{tilt-definition}). \label{haleslaw}}
\end{figure}

Hale's law appears in Figure (\ref{haleslaw}), where filled ``$\bullet$'' and empty ``$\circ$'' symbols represent sunspots in the northern and southern hemispheres, respectively.  In Cycle 23, the leading (trailing) polarities are positive (negative) in the northern hemisphere; the tilt angles are in quadrants  I or IV  ($|\gamma|\leq\pm90^\circ$, c.f.~Figure \ref{tilt-definition}). The leading (trailing) polarities are negative (positive) in the southern hemisphere; the tilt angles are in  quadrants II or III ($90^\circ\leqslant\gamma<180^\circ$ or $-180^\circ\leqslant\gamma <-90^\circ$). The sense of the polarity reverses between the hemispheres in Cycle 24.  Figure  (\ref{haleslaw}) shows that most sunspots obey Hale's law by following this pattern. Exceptions are plotted in black ``$\bullet$'' circled with green.  These  are the ``anti-Hale'' sunspots (hemispheres are not distinguished here).

Except for numbers of sunspots with respective hemispheres and cycles, Table (\ref{thaleslaw}) also shows the statistics of the ``Hale'' and ``anti-Hale'' sunspots for each hemisphere and cycle.  The ``anti-Hale'' sunspots are a stubborn minority: over all, the fraction of ``anti-Hale'' sunspots   is $(8.69\pm0.57)\%$ in Cycle 23, and $(7.11\pm0.64)\%$ in Cycle 24. Within the uncertainties (we quote the Poisson error), both  fractions of ``anti-Hale'' sunspots are equivalent, and consistent with the mean value  $(8.07\pm0.43)\%$. 

\begin{figure}[t]
\includegraphics[width=1.\textwidth]{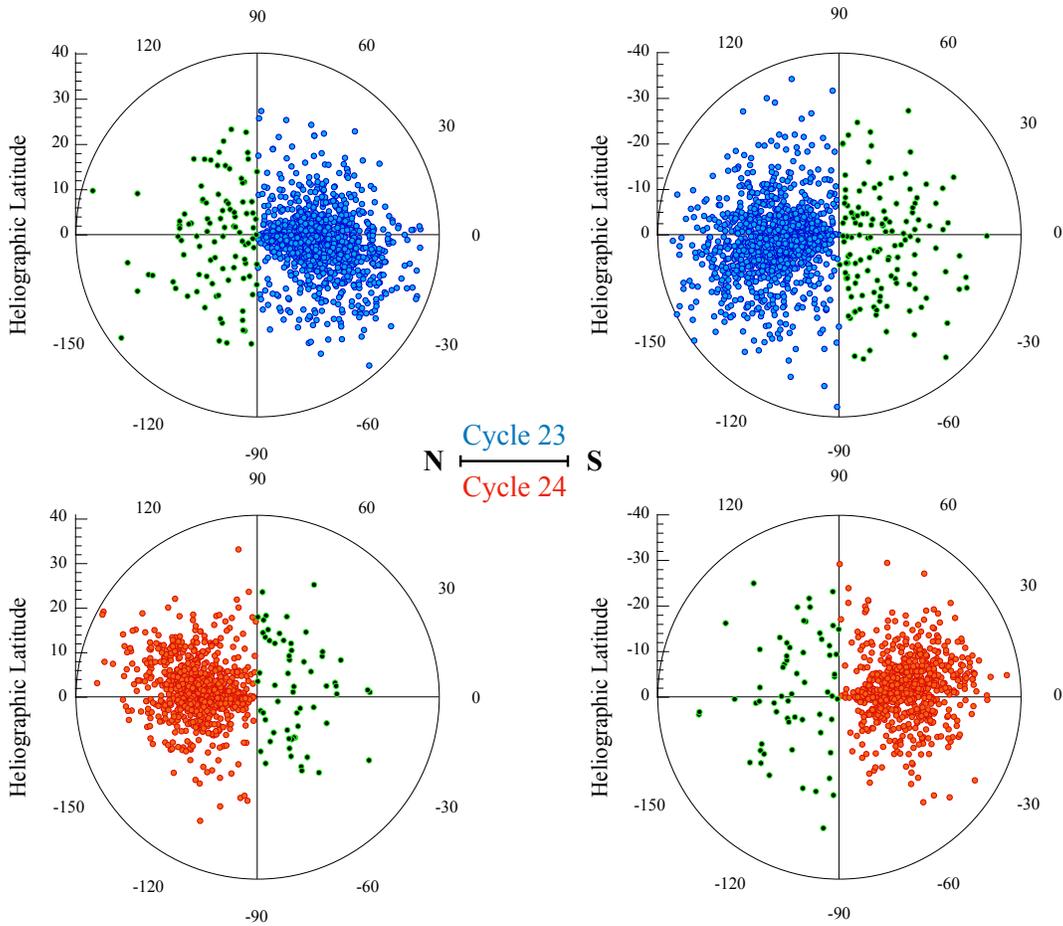}
\caption{Sunspots displayed in polar coordinates, where azimuth shows sunspot tilt angle, and the polar radials give the sunspot latitude. The upper two panels are the sunspots from Cycle 23, and lower two are those from Cycle 24. Left two panels represent the northern hemisphere, and the right two panels represent the southern hemisphere. Black circles with green rings represent ``anti-Hale'' sunspots. \label{polar2tilt}}
\end{figure}

We also  see Hale's law in  polar coordinates, where the azimuthal angle represents the sunspot tilt, and the radius represents the sunspot latitude, from 0\degr~to 40\degr~(Figure \ref{polar2tilt}).  In the Figure, the upper panels show sunspots in Cycles 23 and the lower panels in  Cycle 24. The northern hemisphere spots are plotted in the left two panels and the southern hemisphere spots on the right.   This presentation distinguishes ``Hale'' from ``anti-Hale'' spots particularly clearly:  ``Hale'' sunspots  fall in  quadrants I \& IV or II \& III, depending on hemisphere and cycle while ``anti-Hale'' sunspots (plotted with black dots circled in green) occupy the other two quadrants.  

\subsection{Sunspot Magnetic Flux}

\subsubsection{Cumulative and Average Magnetic Fluxes}
The cumulative  magnetic flux, $\sum\Phi$, is the  magnetic flux integrated over an entire cycle. The average  magnetic flux, $\bar\Phi$, is the  flux averaged over the number of sunspots within a certain category. Both quantities are listed in Table (\ref{tmb}). 

The Table shows that the cumulative magnetic fluxes of the ``Hale'' sunspots are much larger than those of  the ``anti-Hale'' sunspots. For example, in Cycle 23, $\sum\Phi$(Hale)$/\sum\Phi$(anti-Hale)$=14.8\pm1.0$ while in Cycle 24, the ratio is $21.3\pm1.9$.  The average between cycles is  $\sum\Phi$(Hale)$/\sum\Phi$(anti-Hale)$=16.2\pm0.9$. The fraction of the total magnetic flux carried by ``anti-Hale'' sunspots  is $(5.8\pm0.3)\%$, averaged  over the two cycles.

The ratios of average magnetic fluxes in the ``Hale'' and ``anti-Hale'' sunspots are $\bar\Phi$(Hale)$/\bar\Phi$(anti-Hale)$=1.41\pm0.09$ for Cycle 23, and $1.63\pm0.15$ for Cycle 24.   These values are consistent with their weighted mean $\bar\Phi$(Hale)$/\bar\Phi$(anti-Hale)$=1.42\pm0.08$.

The cumulative magnetic fluxes  in Table (\ref{tmb}) dramatically decreased from Cycle 23 to 24, $\sum\Phi(23)/\sum\Phi(24)=2.62\pm0.06$. This reduction is not only due to the decreased number of  sunspots, but also to a reduction in the average magnetic flux per spot. Specifically, the number of sunspots in Cycle 23 is 1.56 times that in Cycle 24 while the average magnetic flux ($\bar\Phi$) in Cycle 23 is $1.68\pm0.04$ times that in Cycle 24. The product of these factors gives a decrease in the cumulative magnetic flux  by a factor 2.6 from Cycle 23 to Cycle 24.



\subsubsection{Magnetic Flux vs. Latitude}
\begin{figure}[t]
\includegraphics[width=1.\textwidth]{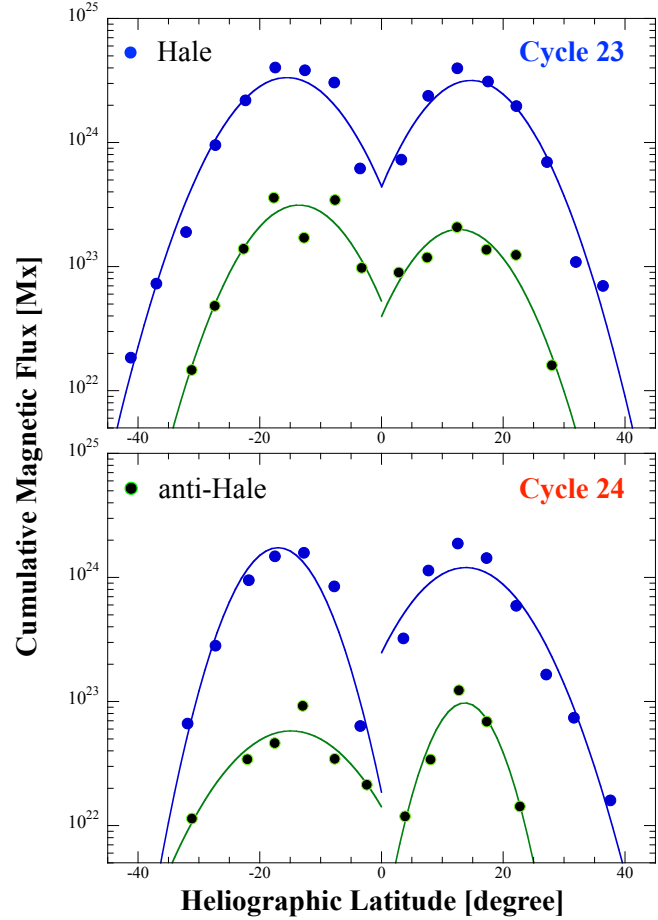}
\caption{Latitude distribution of the cumulative magnetic flux per cycle per ($5^\circ$) latitudinal bin  with ``Hale'' (blue) and ``anti-Hale'' (black/green) populations, respectively. Blue and green curves are parabolas fitted to the fluxes in each hemisphere, to guide the eye.  The upper panel is from Cycle 23, and the lower panel is from Cycle 24. \label{magflux2lat}}
\end{figure}

Figure (\ref{magflux2lat}) shows the  cumulative magnetic flux  distributions with latitude for each hemisphere and cycle, $\sum\Phi(\phi)$. The filled circles represent the integrated magnetic flux over an entire cycle binned by $5^\circ$ in latitude.  ``Hale'' and ``anti-Hale'' sunspots are represented separately by blue and black/green colors.   The  solid lines are parabolas fitted to $\log_{10}(\Phi)~vs.~\phi$, to guide the eye.  The latitude distributions of the flux are remarkably similar between hemispheres and cycles, and between ``Hale'' and ``anti-Hale'' spots.  The latitudes of peak flux are summarised in the last column in Table (\ref{tmb}). The average peak latitude for the emergence of flux is  $14.5^\circ\pm0.5^\circ$, regardless of hemisphere, cycle or ``Hale'' vs.~``anti-Hale'' nature of the spots. 

\begin{figure}[t]
\includegraphics[width=1.\textwidth]{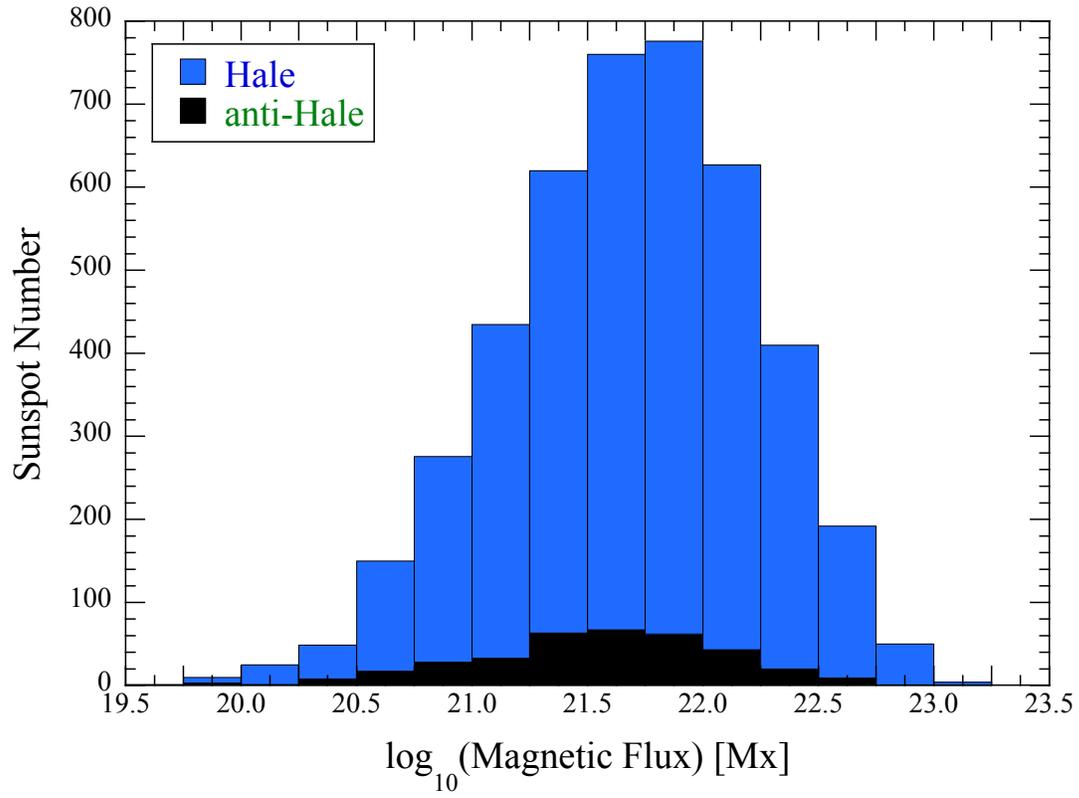}
\caption{Sunspot numbers as a function of $\log_{10}$(magnetic flux). The histograms represent all ``Hale'' (blue) and ``anti-Hale'' (black) sunspots from Cycles 23 and 24. \label{magflux-his}}
\end{figure}

Figure (\ref{magflux-his}) shows the sunspot number distributions vs.~sunspot magnetic fluxes. A striking feature is that both ``Hale'' (blue) and ``anti-Hale'' (black) sunspot populations have similar distribution functions. 

\subsubsection{Magnetic Flux and Pole Separation}
\label{flux_pole}
\begin{figure}[t]
\includegraphics[width=1.\textwidth]{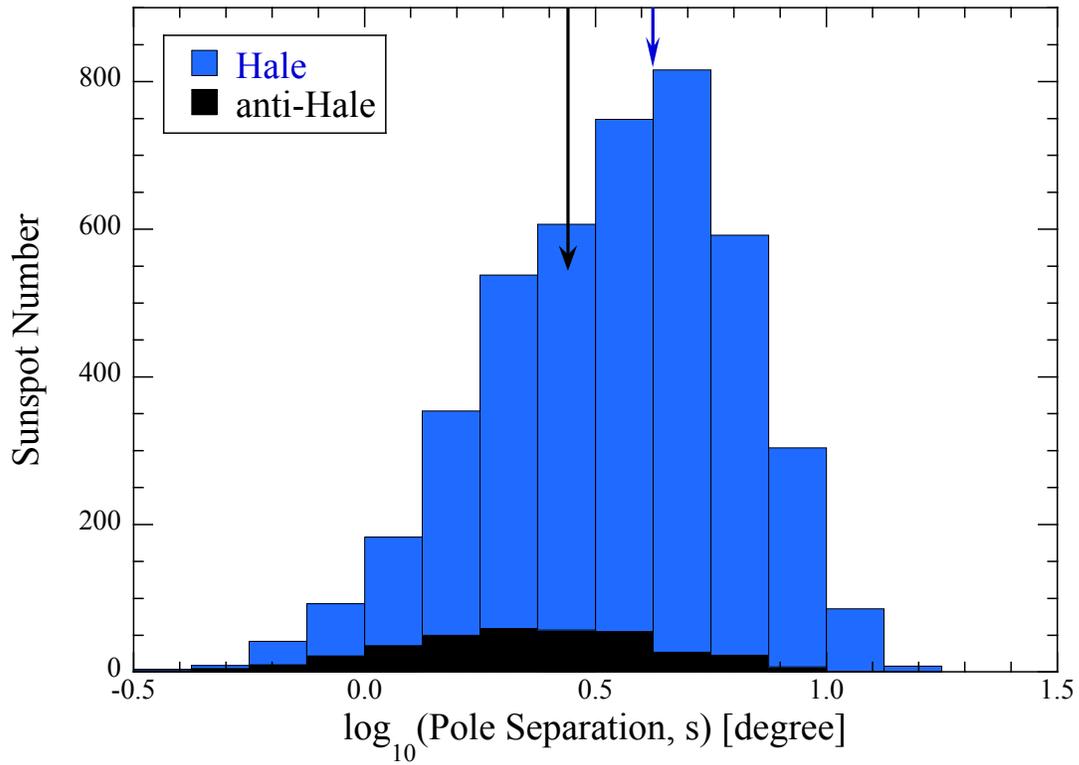}
\caption{Sunspot pole separation distributions. Blue represents the ``Hale'' sunspots, and black represents the ``anti-Hale'' sunspots. Blue and black arrows  represent the logarithmic averaged pole separation for ``Hale'' and ``anti-Hale'' populations, respectively.  They point at $\log_{10}\bar s=0.62$ for ``Hale'' and $\log_{10}\bar s=0.44$ for ``anti-Hale''. \label{histogram-s}}
\end{figure}

We define the sunspot pole separation, $s$\degr, as the distance between the best-fit centers of mass of the positive and negative polarities. The pole separation is given by the spherical cosine law:

\begin{equation}
s=\arccos[\sin \phi_1\sin \phi_2+\cos \phi_1 \cos \phi_2\cos(\Delta \lambda)]\times(180\degr/\pi)
\end{equation}

\noindent where $\phi_1$ and $\phi_2$ are the heliographic latitudes of the two poles  and $\Delta \lambda$ is the difference between the heliographic longitudes of the poles. Table (\ref{tflux2s}) gives the logarithmic average pole separations $\log_{10} (s) = 0.62\pm0.01$  for ``Hale'' and  $\log_{10} (s) = 0.44\pm0.02$ for ``anti-Hale'' sunspots, respectively.  The difference is statistically significant, with the average pole separation of ``Hale'' sunspots being larger than that of ``anti-Hale'' sunspots. This is also seen in the histogram of Fig. (\ref{histogram-s}), indicated  by two arrows.  

\begin{figure}[t]
\includegraphics[width=1.\textwidth]{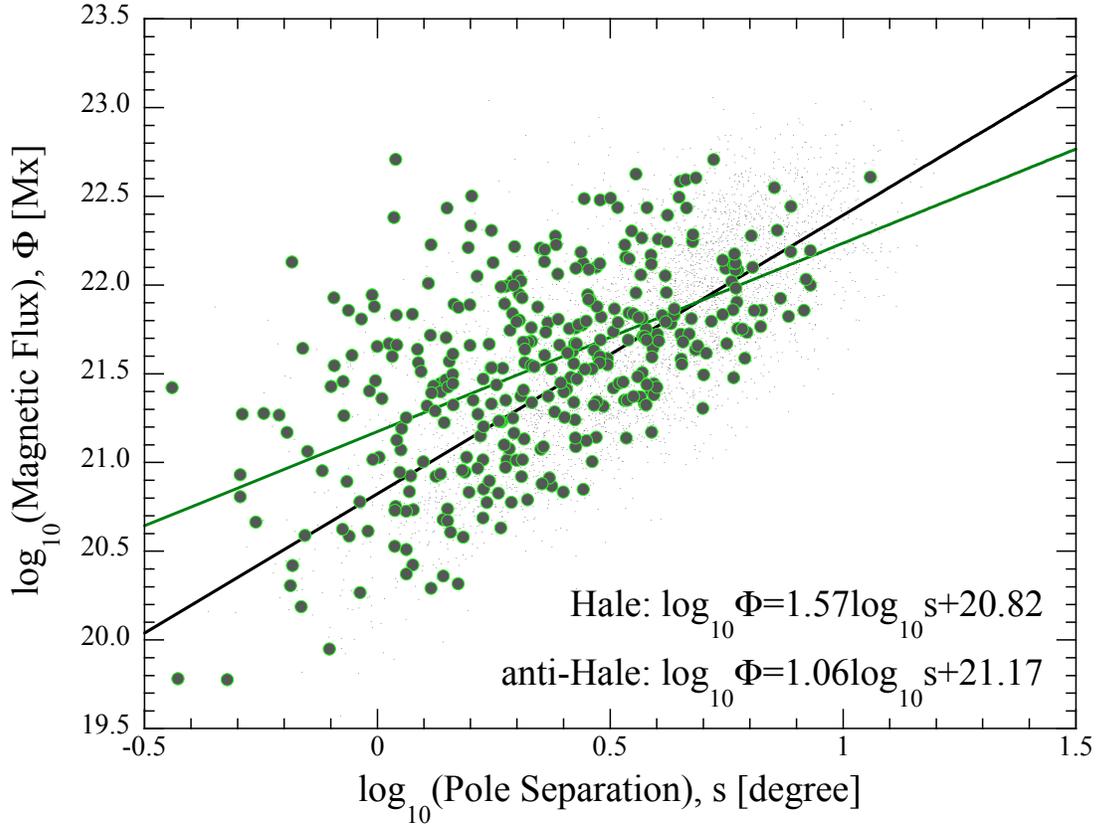}
\caption{Sunspot magnetic flux ($\Phi$) versus polarity pole separation ($s$) on a log-log scale. Hale sunspots are plotted with small black dots; and  ``anti-Hale'' sunspots are plotted ``$\bullet$'' circled in green rings. Grey and green straight lines show linear fits to ``Hale'' and ``anti-Hale'' sunspots, respectively. The fitting parameters and errors are also found in Table (\ref{eflux2s}).  \label{flux2s}}
\end{figure}

In Figure  (\ref{flux2s}), the sunspot magnetic flux ($\Phi$) is plotted as a function of the pole separations ($s$), showing that these quantities are clearly correlated.  Pearson correlation coefficients between the two parameters  are 0.75 for the ``Hale'' and 0.54 for the ``anti-Hale'' populations, both significant with  the $P < 10^{-5}$ probability.   We fitted a relation of the form

\begin{equation}
\log_{10} \Phi(s)=k_s\log_{10} s +C_s
\label{eflux2s}
\end{equation}

\noindent where $k_s$ and $C_s$ are constants for both  ``Hale'' and ``anti-Hale'' populations, using data from both cycles and hemispheres.  The fitting parameters, their uncertainties, and the correlation coefficients are listed in Table (\ref{tflux2s}). The regression lines are plotted in Figure (\ref{flux2s}) in grey for the ``Hale'' sunspots, and green for ``anti-Hale'' sunspots.   The dependence of the sunspot magnetic flux on pole separation, $s$, is different for ``Hale'' and ``anti-Hale'' populations. Expressed as power laws, we find $\Phi(s)=(6.61^{+0.47}_{-0.44})\times10^{20}s^{1.57\pm0.02}$ for ``Hale'' sunspots and $\Phi(s)=(14.79^{+4.71}_{-3.57})\times10^{20}s^{1.06\pm0.09}$ for ``anti-Hale'' sunspots. 

The relationship between magnetic flux and pole separation was examined by \citet{1989SoPh..124...81W} using Kitt Peak magnetograms. They obtained $\Phi(s)=4\times 10^{20}$  $s^{1.3}$ without distinguishing ``Hale'' from ``anti-Hale'' populations. To compare with Wang's work, we fitted Equation (\ref{eflux2s}) to all sunspots (c.f. last row in Table \ref{tflux2s}), to find $\Phi(s)=7.59^{+0.34}_{-0.36}\times 10^{20} s^{1.49\pm0.02}$.  This result is consistent with that obtained by \citet{1989SoPh..124...81W} except that our multiplicative constant, $\Phi_0=7.59\times 10^{20}$, is about twice their value of $4.0\times 10^{20}$.  This factor of two occurs  simply because Wang and Sheeley described the single polarity flux while we present the sum of the absolute values of the north and south components.  Our estimate of the power index, $s$ = 1.49$\pm$0.02, is slightly larger than $s$ =1.3 in Wang and Sheeley but this difference is probably within the uncertainty of their determination (Y.-M. Wang, private communication).


\subsection{Sunspot Magnetic Tilt Angles}
The sunspot tilt angle, $\gamma$, is the angle between the magnetic axis and the Sun's azimuthal direction or equator. The tilt angle range is [$-90^\circ, 90^\circ$]. We define tilt angles positive when the sunspot magnetic axes tilt toward to the equator, negative when the axes tilt away from the equator. 

\subsubsection{Tilt Angle Statistics}
The statistics of sunspot tilt angles are summarized in Table (\ref{ttiltangles}). We list  the average ($\bar\phi$), median ([$\phi$]) and the standard deviation ($\sigma_\phi$) for sunspot latitudes; and the average ($\bar\gamma$), median ([$\gamma$]) and the standard deviation ($\sigma_\gamma$) for sunspot tilt angles.  ``Hale'' and ``anti-Hale'' populations are presented separately for each hemisphere and cycle. 

The average and median sunspot latitudes are roughly identical between the ``Hale'' and ``anti-Hale'' populations, hemispheres and cycles. We obtain the average latitude of all sunspots, $\bar\phi=\pm(15.55^\circ\pm0.23^\circ)$. The listed uncertainty is the Poisson error. 

The average absolute tilt angle determined from all sunspots is $\bar\gamma=4.58^\circ\pm0.07^\circ$. This is similar to $4.2^\circ\pm0.2$\degr~obtained by \citet{1991SoPh..136..251H}. For ``Hale'' sunspots, we find  average tilt angle $\bar\gamma=5.49\degr\pm0.09\degr$. For ``anti-Hale'' sunspots, $\bar\gamma=-5.84\degr\pm0.31\degr$. In general, the ``Hale'' sunspot magnetic axes tilt toward the equator, but the ``anti-Hale'' axes tilt away from it.

\subsubsection{Joy's Law}
In the polar plots of Figure (\ref{polar2tilt}), ``Hale'' sunspots concentrate near the horizontal axes, and appear in broad, fan-like clusters. This is because the tilt angles of Hale's sunspots generally increase with increasing latitude, following Joy's law. The ``anti-Hale'' sunspots are scattered in the  quadrants not occupied by ``Hale'' sunspots.  Their distributions show no dependence on the latitudes.  Here, we calculate best-fit values of Joy's law parameters using only the ``Hale'' sunspots.  

Slightly different formulations of Joy's Law are used in the literature.  To compare with these different formulations, we fit the following three  functions to the tilt vs.~latitude data.  
First, to compare with the effect of  Coriolis force, which varies in proportion to $\sin(\phi)$, we examine Joy's law written as \citep{1989SoPh..124...81W}, 

\begin{equation}
\sin\gamma = k_J\sin \phi+C_J.
\label{joysin}
\end{equation}

\noindent Second, we fitted the simple form \citep{2012ApJ...758..115L}:

\begin{equation}
\gamma=k_\phi\phi+C_\phi.
\label{joydirect}
\end{equation}

\noindent Finally, we fitted \citep{2012ApJ...745..129S}

\begin{equation}
\gamma=k_\circ\sin \phi + C_\circ.
\label{joystenflo}
\end{equation}


In these equations,  $k_J$, $k_{\phi}$ and $k_\circ$ are  the Joy's slopes ($k_\phi$ and $k_\circ$ have units of degree of tilt per  degree of latitude) and $C_J$, $C_{\phi}$ and $C_0$ are the equatorial ($\phi=0\degr$) tilt angle  (``Joy's constant''). 
Results of the fits are listed in Table (\ref{tjoyslaw}), for each hemisphere and  activity cycle.  Note that, traditionally, Joy's law is derived from a few latitude bins to increase the signal-to-noise. But with our large data set, we derive Joy's law directly.  

Table (\ref{tjoyslaw}) summarises the parameters for Equations (\ref{joysin}), (\ref{joydirect}) and (\ref{joystenflo}). In all cases the Joy's constant intercepts, $C_J$, $C_{\phi}$ and $C_\circ$, are  statistically consistent with zero.  Furthermore, within the uncertainties,  the derived  parameters are independent of hemisphere and cycle number, other than for the expected sign differences. 

\begin{figure}[t]
\begin{center}
\includegraphics[width=0.5\textwidth]{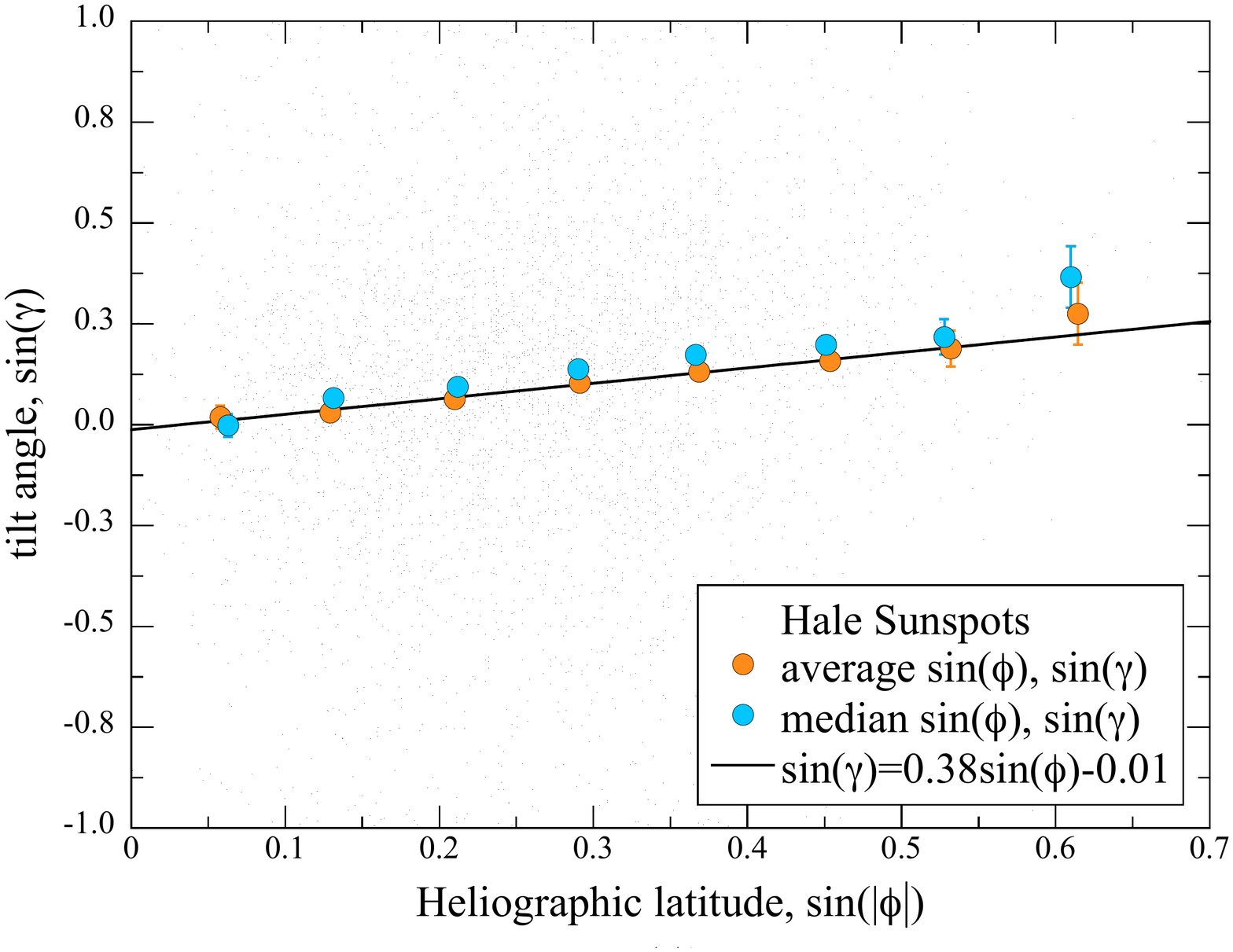}
\includegraphics[width=0.5\textwidth]{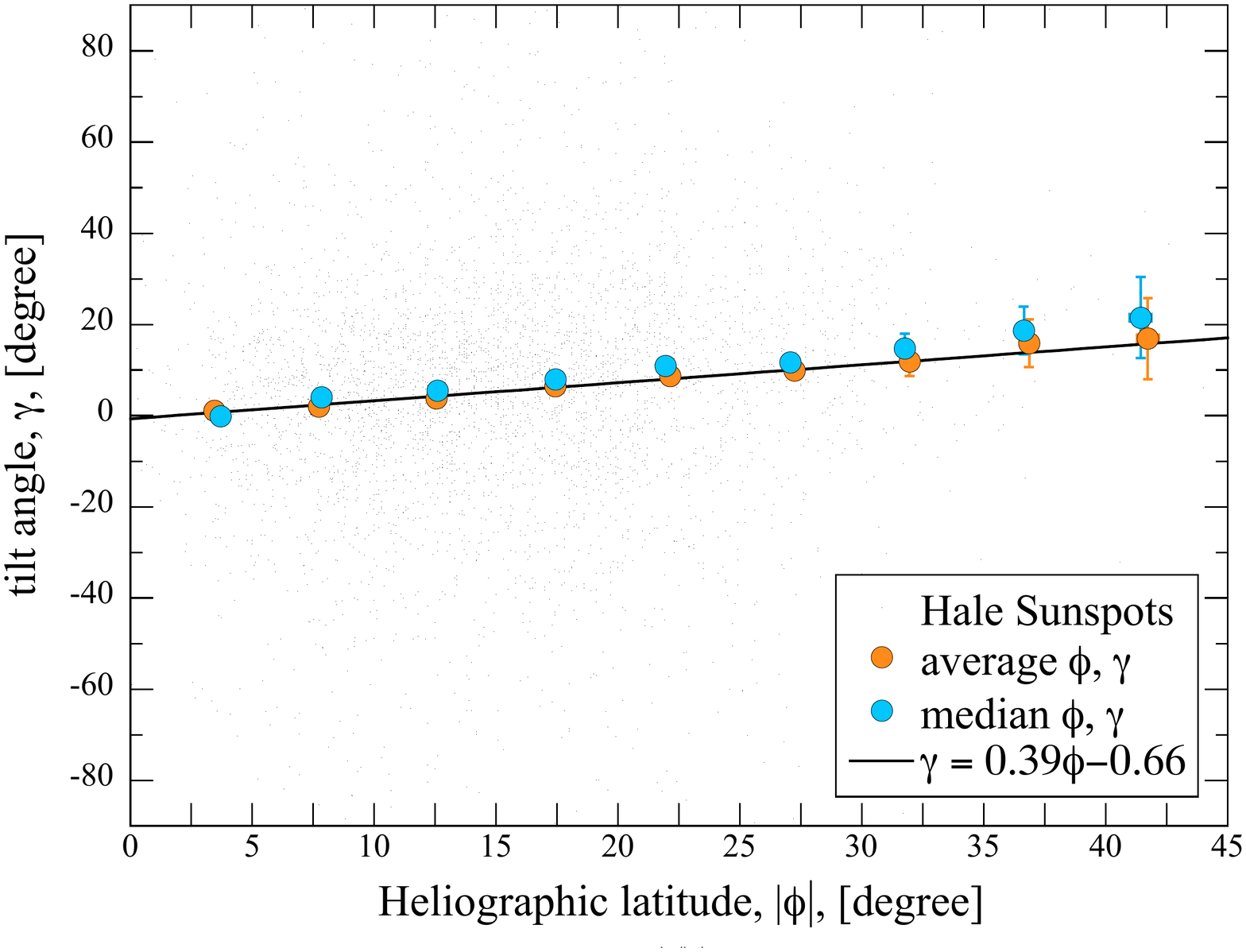}
\includegraphics[width=0.5\textwidth]{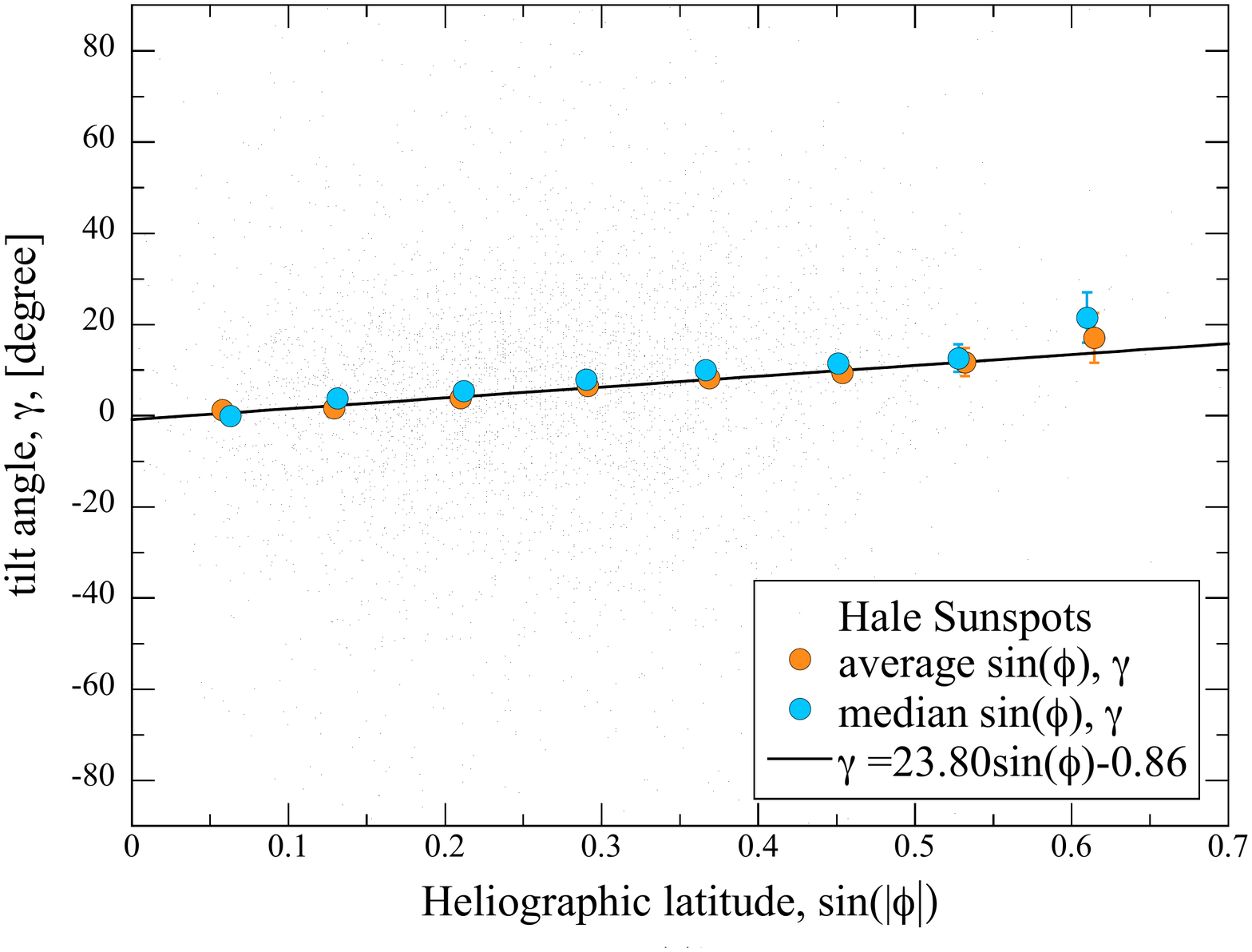}
\caption{Joy's law: Tilt angles ($\gamma$) as a function of absolute heliographic latitude ($|\phi|$). Panels represent three expressions of Joy's law: Equations (\ref{joysin}), (\ref{joydirect}) and (\ref{joystenflo}), which are linear least-square fitted to all ``Hale'' sunspots (see straight lines). The blue and orange filled circles represent average and median values of sunspot $|\phi|$ and $\gamma$ binned by $5^\circ$; and $\sin|\phi|$ and $\sin\gamma$ binned by 0.08. These data points show the trends of tilt angle as function of latitude, but are not used to derive  Joy's laws. The error bars in both $\phi$ and $\gamma$ directions are the standard deviations of mean. \label{fjoyslaw}}
\end{center}
\end{figure}

Having established that there are no significant differences between hemispheres or solar cycle numbers, we  obtain  Joy's law expressions using tilt angle determinations for all ``Hale'' sunspots as a function of absolute latitude, $|\phi|$ (i.e.~we merge the data for both hemispheres and cycles).  The resulting  three forms of Joy's law are plotted in Figure (\ref{fjoyslaw}).  All ``Hale'' sunspots are plotted with ``$\cdot$'' symbols, and we show absolute latitudes, $|\phi|$. The red and the blue filled circles represent the average, median latitudes and tilt angles every $5^\circ$ (0.0833 for $\sin]phi$) of the (sine) latitudinal bin, while the vertical and horizontal error bars are the standard deviations of the means. These binned points are to guide the eye, but are not used for fitting the Joy's law parameters.  

From Equation (\ref{joysin}), we find $\sin\gamma$ =(0.38$\pm$0.05)$\sin\phi$ -(0.01$\pm$0.02).  This Joy's slope is slightly but not significantly smaller than 0.48 derived earlier by fitting the flux-weighted tilt angles \citep{1991ApJ...375..761W}. From Equation (\ref{joydirect}), we find $\gamma=(0.39\pm0.06)\phi-(0.66\pm1.00)$.  Joy's slope here is consistent with $0.5\pm0.2$ obtained in our previous study \citep{2012ApJ...758..115L}, and the current work offers a more accurate measurement. The Joy's slopes differ insignificantly between Equations (\ref{joysin}) and (\ref{joydirect}) because, for small sunspot tilt angles, $\gamma<30^\circ$, and latitudes, $\phi<20^\circ$, $\phi_r\sim\sin\phi$, and $\gamma_r\sim\sin\gamma$, where $\phi_r$ and $\gamma_r$ are expressed in radians. Finally, we find $\gamma$ =(23.80$\pm$3.51)$\sin\phi$ - (0.86$\pm$1.03) from Equation (\ref{joystenflo}).  Our determination of Joy's slope in Equation (\ref{joystenflo}) is $\sim$2.5$\sigma$ smaller than $32.1^\circ\pm0.7^\circ$ as obtained by \citet{2012ApJ...745..129S}. If this difference is real it could be due to the inclusion of bipolar regions of all sizes, from ephemeral to large sunspots,  in the study of \citet{2012ApJ...745..129S}. In all three derived forms of Joy's law, the intercepts are statistically consistant with ``0''.

\subsubsection{Sunspot Tilt Angle vs~Pole Separation}
\begin{figure}[t]
\includegraphics[width=1.\textwidth]{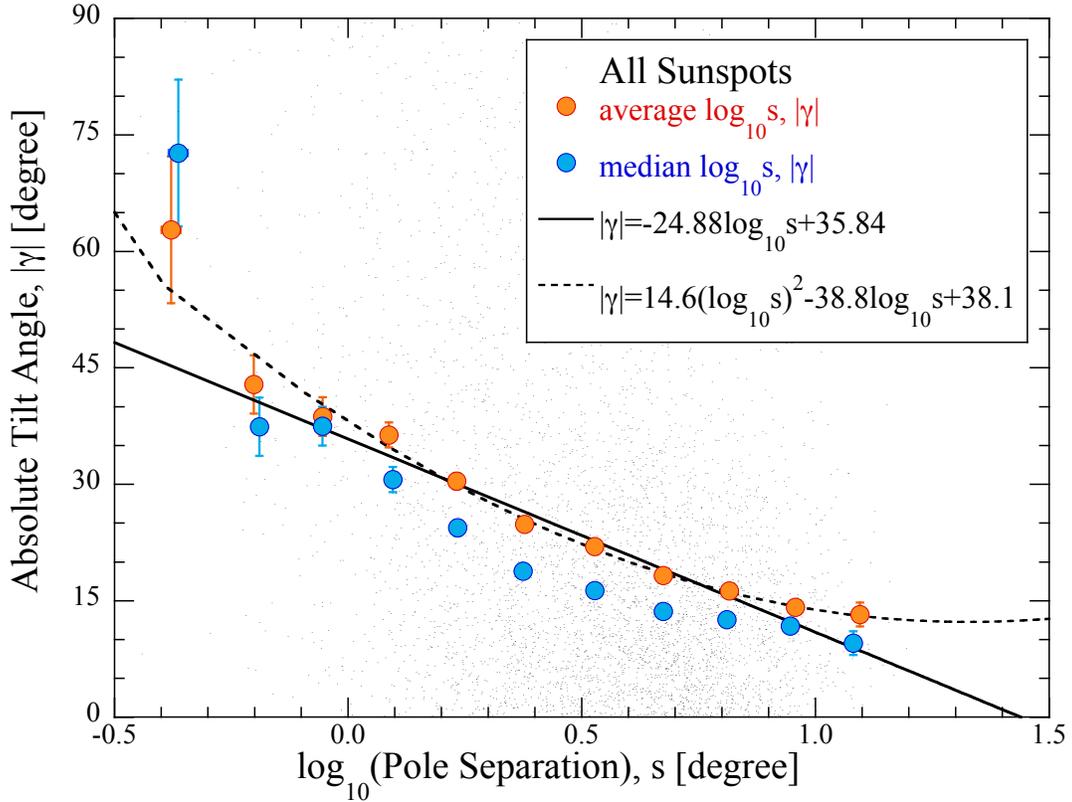}
\caption{Tilt Angle as a function $\log_{10}$(pole separation). All sunspots are plotted with ``$\cdot$'' symbols. Orange and blue filled circles represent the average and median data points over a $\sim0.15$ logarithmic bin pole separation. They show the trend, but are not used to fit the equation. The solid line shows Equation (\ref{etilt2s}) fitted to all sunspots. The dashed curve is a parabola fitted to all sunspots. The error bars are the standard deviation of mean. \label{ftilt2s}}
\end{figure}

The Pearson's correlation coefficient between absolute tilt angle, $|\gamma|$, and pole separation, $\log_{10}(s)$, is -0.33 for all sunspots. This is with  $P<10^{-5}$ at 0.01 confidence level implicating that the two parameters are highly correlated.  Fig (\ref{ftilt2s}) plots all sunspots with black dots. To guide the eye, the orange and light blue filled circles represent the average and median data points at the 0.15 $\log s$ bin width.  For a crude estimate, we fit the following equation to all sunspot data:
\begin{equation}
|\gamma|=k_\gamma log_{10}s+C_\gamma
\label{etilt2s}
\end{equation}


\noindent  Unlike the relations shown in Figure (\ref{flux2s}), the sunspots are much more scattered in Figure (\ref{ftilt2s}) (see ``$\cdot$'' symbols). A linear fit to all sunspots gives  $|\gamma|=(-24.88\pm1.08)\log s+(35.84\pm0.64)$. This is plotted with a solid line in the Figure (\ref{flux2s}). A visual inspection of  the discrete data points inspires a parabola function. A least square fit to all sunspots gives $|\gamma|=(14.6\pm3.1)(\log s)^2-(38.8\pm3.1)\log s+(38.1\pm0.8)$ (see the dashed curve). The relation shows that the tilt angle decreases with increasing pole separation. 


\citet{1993SoPh..145..105H} used  daily white-light photographs taken at Mt. Wilson to reach a different conclusion.  A linear least-square fit to their data gives a slope of 0.058 deg Mm$^{-1}$ indicating a positive correlation between tilt angle and pole separation. To compare with Howard's measurement we recomputed the fit between $|\gamma|$ and $s$ expressed in meters (instead of degrees) for {\it all  sunspots}.  The linear fit gives the slope $-0.20\pm 0.01$ deg Mm$^{-1}$, which is very different from 0.058 deg Mm$^{-1}$ and opposite in sign. We speculate that the discrepancy may be due to the use of magnetograms in our study versus  white light photographs in Howard's work. For example, magnetograms inevitably draw in surrounding plage regions but the white light photographs might have included only sunspots and their penumbrae.  In support of this possibility, and of the current work, we note that  an independent but smaller study of 203  active regions using magnetograms from Huairou Solar Observatory also found that the sunspot tilt angles decrease with increasing pole separation \citep{1999SoPh..189..305T}.

\section{Discussion}
As remarked above and shown in Table \ref{thaleslaw}, the fraction of ``anti-Hale'' sunspots (namely $8.07\%\pm0.43$\%), is fixed  with respect to hemisphere and cycle number. This fraction is consistent with previous studies in which the ``anti-Hale'' fraction ranges from a few to $\sim$10\% \citep{1989SoPh..124...81W,2009ARep...53..281K,2012ApJ...745..129S,2012ApJ...758..115L,2014ApJ...797..130M,2015MNRAS.451.1522S}.  It is most consistent with ($8.4\pm0.8$)\% obtained by \citet{2014ApJ...797..130M} and ($8.2\pm0.3$)\% from our previous work \citep{2012ApJ...758..115L}. This is probably because these works employ the same magnetic field data.

It is not clear why one out of twelve sunspots should persistently violate the hemispheric polarity rule. The evidence shows that many of these sunspots are not evolved from the ``Hale'' population, but have irregular polarity arrangements in the beginning, as was also noted by \citet{2012ApJ...745..129S}. The question is how sunspots emerge with irregular magnetic polarity orientations, with the toroidal fields pointing in the opposite direction. In numerical simulations, ``anti-Hale'' sunspots are the result of either weak magnetic fields, or  ``hemisphere crossing due to convective flows'' \citep{2013SoPh..287..239W}. Indeed, the cumulative magnetic fluxes of ``anti-Hale'' sunspots are at the level of a few percentage ($5.8\pm0.3$)\%  of total magnetic flux; ; the average magnetic fluxes of ``anti-Hale'' sunspots are also smaller, only $\sim70$\% of those of the ``Hale'' population. However, some individual ``anti-Hale'' sunspots do possess large magnetic fluxes, as evident in Figure (\ref{magflux-his}).  Table (\ref{ttiltangles}) shows that the average and median latitudes of ``anti-Hale'' sunspots do. not differ significantly from ``Hale'' population . It is yet to be examined what special properties are possessed by ``anti-Hale'' sunspots, and what roles they play in the solar cycle progression.

The ``Hale'' sunspots can be further divided into two sub-populations, namely those with  magnetic axes tilted towards the equator and those tilted away from it.  The former are labeled ``Hale/normal'' and the latter ``Hale/inverted'' by \citet{1989SoPh..124...81W}.  \citet{2015ApJ...808L..28J} and \citet{2018ApJ...863..116W} attribute the weakness of solar cycle 24 to a few large ``Hale/inverted'' sunspot groups appearing near the equator in Cycle 23. Their argument is that the following polarities of these sunspots traverse the equator to eventually weaken the global dipole. In reality, it is observed that Hale/normal and Hale/inverted spots are continuous states of an evolving sunspot group. Statistically, more than half of sunspots are  Hale/normal, and more than 1/3 sunspots are  Hale/inverted.  Table (\ref{haless}) shows that the fractions of Hale/normal spots are $\sim$55\% in Cycle 23 and $>60\%$ in Cycle 24, a modest difference.  The Table also shows that the average latitudes of Hale/inverted spots are lower than Hale/normal spots by $\sim1.5^\circ$.  However, in view of the uncertainties in the measurements, and because we are comparing data from only two sunspot cycles, it is not clear that these differences  are sufficient to cause the prolonged minimum in the end of Cycle 23. More data are needed to be able to address this issue with greater confidence.

Joy's law in  ``Hale'' sunspots can be explained by considering the action of the Coriolis force on the flow directions in the flux tubes \citep{2011ApJ...741...11W}, or on the expansion/contraction of the flux tubes \citep{1991SoPh..132..257H,1991SoPh..136..251H,1991ApJ...375..761W}.  The magnitude of the tilts is determined by the time of emergence, the solar rotation rate (the latitude) (see Equation \ref{coriolis}). Statistically, the average emergence time $\Delta t\sim1.8$ [days] which is from sunspots appearing in the surface to their magnetic flux reaching maximum; the rotation period is 27.3 [days] for the average sunspot latitude $15.55^\circ$; therefore, $\sin\gamma \sim 0.4\sin\phi$ by Equation (\ref{coriolis}). This agrees with the currently measured Joy's law $\sin\gamma=(0.38\pm0.05)\sin\phi$. A similar calculation by \citet{1991ApJ...375..761W} results in $\sin\gamma\sim 0.5\sin\phi$ with the slightly larger coefficient resulting from the use of a slight longer emergence time, $\Delta t=2.2$ days. Over all, our observations is consistent with the hypothesis that Joy's law is produced by the Coriolis force acting on emerging sunspot magnetic flux tubes.

It is inevitable that the determination of sunspot tilt angles is dependent of the area sizes of sunspots. An example is a discrepancy between Joy's law obtained from white-light images and from magnetograms. In general, the Joy's slope is lower  from the white light images, $k_\phi \sim 0.2$- $0.3$ \citep{1919ApJ....49..153H,1991SoPh..136..251H,2010A&A...518A...7D,2012Ge&Ae..52..999I,2013SoPh..287..215M,2018arXiv180410479I} than from the magnetograms, $k_\phi\sim 0.5$ \citep{1989SoPh..124...81W, 2012ApJ...745..129S,2012ApJ...758..115L}. \citet{2015ApJ...798...50W} attribute the discrepancy to the inclusion of plage areas in the measurements with magnetograms. The plage regions are not seen in white light images.  The same conclusion is also drawn by \citet{1996SoPh..167...95H,1996SoPh..169..293H} using Mt Wilson data who noticed plages normally have higher tilt angles than those of sunspot groups. Our observations support this assessment.  The average magnetic strength of active regions is at a magnitude of a few hundreds gauss, which are the signature of plages (see Fig. \ref{bfield}). Whether or not the Joy's slope discrepancy is caused by the sizes of sunspot regions is worth  investigation in  subsequent work. 

Our finding, that Joy's law does not vary  with hemisphere or with solar cycle,  is different from a recent study by \citet{2018arXiv180707913T}.  These authors studied sunspot cycles 15 to 24 using  white-light sunspot drawings. First, they find that tilt angles reach a peak at 20\degr-30\degr~latitudes then decrease towards the poles. This local maximum is not present in our data. Their observation results in Joy's formula $\gamma$=(0.20$\pm$0.08)$\sin(2.80\phi)$+(-0.00$\pm$0.06), with a frequency number for the latitude which is quite different from other studies on Joy's law. Second,  they find that the odd and even cycles have different latitudinal tilt angle profiles. Again, the fact that their conclusion differs from the current work might be partly attributed to the use of sunspot drawings as opposed to magnetograms. However, when they derived  Joy's law in the form of Equation (\ref{joydirect}) the result, $\gamma$=(0.41$\pm$0.18)$\phi$+(0.00$\pm$0.06), is consistent with the value derived here, namely $k_{\phi}$ = 0.39$\pm$0.06, with a larger error bar.



The average pole separation is $\log_{10}(s^\circ) = 0.62\pm0.01$ for ``Hale'' sunspots and  $\log_{10}(s^\circ) = 0.44\pm0.02$ for ``anti-Hale'' spots (see Table \ref{tflux2s}). The difference between two sunspot populations is statistically significant. According to \citet{1993A&A...272..621D},  smaller pole separations represent larger magnetic tension, which opposes magnetic buoyancy and increases the rising time. This may explain why ``anti-Hale'' sunspots do not follow  Joy's law because the magnetic tension is the dominant force over Coriolis force within these sunspot magnetic flux systems.

The sunspot magnetic flux increases with increasing pole separation (Section \ref{flux_pole}, see also \citep{1989SoPh..124...81W, 1999SoPh..189..305T}). A new result from the current work is that ``Hale'' and ``anti-Hale'' sunspots behave differently in the relationship between two parameters.  The sunspot magnetic fluxes increase exponentially with pole separation for ``Hale'' sunspot population ($\Phi(s) \propto s^{1.57}$) . But the magnetic fluxes are almost linearly correlated to the pole separation for ``anti-Hale'' sunspot population ($\Phi(s) \propto  s^{1.06}$). This seems to suggest that the topologies of magnetic flux tubes may differ between two populations since the sunspot magnetic flux is dependent of pole separation or vice versa. 

The current study shows that sunspot tilt angles ($\gamma$), pole separations ($s$) and magnetic fluxes ($\Phi$) are interconnected.  An empirical relation between magnetic fluxes and tilt angles can be derived from general relations that we derived earlier: $\log\Phi(s)=1.49\log s+20.88$ and $|\gamma|=-24.88\log s+35.84$. It is $\Phi(\gamma)=1.1\times10^{23}\times(0.87)^{|\gamma|}$. The formulae indicates that sunspot magnetic flux is weakly anti-correlated with the sunspot tilt angles. This is the result of the correlations between $\Phi$, $s$ and $\gamma$ shown by Equations (\ref{eflux2s}) and (\ref{etilt2s}). 

Our observation shows general trends of sunspot magnetic flux. Larger magnetic flux correlates with bigger pole separation and a smaller tilt angle.  Statistically, systems with small magnetic flux have smaller pole separations but larger tilt angles. This is consistent with the general impression \citep{1993A&A...272..621D}, but seems at odds with some observations and simulations \citep{1993SoPh..145..105H,1995ApJ...438..463F,2011ApJ...741...11W,2013SoPh..287..239W}.  As it is shown in Equation (\ref{coriolis}) and discussed by \citet{2015ApJ...798...50W},  tilt angles depend on the buoyant rise time. Large sunspots tend to have strong fields, and rise quickly through the convection zone. They are less likely to be affected by Coriolis force.  On the other hand, the relations demonstrated in this work are based on the surface magnetic field data, and the statistical behavior of sunspots.  More clues about  flux loop emergence can be expected from a study of the time variation of these parameters. 


%

\section{Summary}
We conducted a program of systematic measurements of sunspot parameters using a uniform, high quality dataset including 4385 sunspots erupted in Cycles 23 and 24. We measured magnetic tilt angles,  fluxes, areas and pole separations using full-disk magnetograms taken by MDI/SoHO and HMI/SDO. 
We used this dataset to characterize differences between the  ``Hale'' and ``anti-Hale'' sunspot populations.  

Our results about ``Hale'' and ``anti-Hale'' sunspot populations are:

\begin{enumerate}

\item The ``anti-Hale'' sunspots constitute ($8.1\pm0.4$)\% of all sunspots.  This fraction is constant with respect to hemisphere and solar cycle number.

\item The ``Hale'' and ``anti-Hale'' populations have similar latitudinal distributions (with mean value $\bar \phi=15.6^\circ\pm0.2^\circ$) but differ in the distributions of tilt angles and magnetic fluxes.    The average tilt angle of the ``Hale'' spots is $\bar\gamma=5.49\degr\pm0.09\degr$ and of the ``anti-Hale'' spots  $\bar\gamma=-5.84\degr\pm0.31\degr$.   On average, the ``Hale'' sunspots carry a cumulative magnetic flux 16 times that of ``anti-Hale'' sunspots, and have an average magnetic flux per sunspot 1.4 times that of ``anti-Hale'' sunspots. 

\item The average pole separation of ``Hale'' spots, $4.18^\circ\pm0.07\degr$, is larger than that of ``anti-Hale'' spots, $2.74^\circ\pm0.15\degr$. This suggests that ``anti-Hale'' sunspot magnetic flux loops have generally stronger magnetic tension than ``Hale'' sunspots do.

\item Joy's law for ``Hale'' sunspots is equally well-described by 1) $\sin\gamma=(0.38\pm0.05)\sin\phi)$, 2) $\gamma=(0.39\pm0.06)\phi$, and 3) $\gamma=(23.8\pm3.5)\sin\phi$.

\item Empirically, we find the sunspot magnetic flux as a function of pole separation, $\Phi(s)= (6.6^{+0.5}_{-0.4})\times 10^{20}s^{1.57\pm0.02}$ for ``Hale'' sunspot populations, and $\Phi(s)=(14.8^{+4.7}_{-3.6})\times 10^{20}s^{1.06\pm0.09}$ for ``anti-Hale'' sunspot populations. For all sunspots, the formula is $\Phi(s)= (7.6^{+0.3}_{-0.4})\times 10^{20}s^{1.49\pm0.02}$.

\end{enumerate}

Other significant results are:

\begin{enumerate}
\item Sunspot emergence is hemispheric asymmetric, with the southern hemisphere dominant in Cycle 23 and the northern hemisphere dominant in Cycle 24. But sunspot eruptions are hemisphere-sysmmetric over a magnetic cycle ($\sim22$ years), i.e. $N_n(23)N_s(24)/N_s(23)N_n(24)\sim 1.04$, is  close to unity.

\item The number of sunspots erupted in Cycle 23 is $\sim$1.6 times that of Cycle 24. The magnetic flux erupted through Cycle 23 is $\sim$2.6 times that of Cycle 24.

\item Statistically, we find the  tilt angles decrease with increasing pole separations according to $|\gamma(s)|=(-24.88\pm1.08)\log_{10}s+(35.84\pm0.64)$.  For a more accurate formula, we fit a parabola: $|\gamma|=(14.6\pm3.1)(\log s)^2-(38.8\pm3.1)\log s+(38.1\pm0.8)$.

\item Empirically, magnetic flux is related to the tilt angle: $\Phi(\gamma)=1.1\times10^{23}\times (0.87)^\gamma$, where $\gamma\geq0$. Sunspot magnetic flux decreases modestly with increasing tilt angle. 

\end{enumerate}

\acknowledgments
JL thanks David Jewitt and Xiaodong Zhou for discussions. She would like to thank Andr\'{e}s Mun\~{n}oz-Jaramillot for discussions and suggestions at the Fall AGU 2015, Mei Zhang for her visit and discussions, and the enthusiastic supports from Drs. Roger Ulrich, Marco Velli, and Aimee Norton. Dr. Y.-M. Wang has read the manuscripts, and provide valuable comments. The comments by the anonymous referee have helped to greatly improve the paper. This work is partially supported by a NASA grant NNX15AF29G awarded to JL. 

%
%

\begin{deluxetable}{ccrrlcrl}
\tablewidth{0pt}
\tablecaption{Hale's Law\label{thaleslaw}}
\tablehead{
\colhead{Cycle\tablenotemark{(a)}} &
\colhead{Hem.\tablenotemark{(b)}} &
\colhead{\#\tablenotemark{(c)}} &
\multicolumn{2}{c}{Hale\tablenotemark{(d)}} &
\colhead{} &
\multicolumn{2}{c}{anti-Hale\tablenotemark{(e)}} \\
\cline{4-5}  \cline{7-8}\\
\colhead{} &
\colhead{} &
\colhead{} &
\colhead{\#} &
\colhead{\%} &
\colhead{} &
\colhead{\# } &
\colhead{\%} 
}
\startdata
23 & N & 1213& 1115 & $91.9\pm2.8$ && 98& $8.1\pm0.8$ \\
& S&  1457 &1323 &$90.7\pm2.5$ &&134 &$9.2\pm0.8$\\
 \hline
  24 & N & 917 &857 &$93.5\pm3.2$&  &60 &$6.5\pm0.8$ \\
 & S& 798 &736 & $92.2\pm3.4$ &&62 & $7.8\pm1.0$ \\
 \enddata
 \tablenotetext{(a)}{Solar cycle number}
 \tablenotetext{(b)}{Hemisphere letters}
 \tablenotetext{(c)}{Number of sunspots}
 \tablenotetext{(d)}{Number,  percentage and its uncertainty of ``Hale'' sunspots}
 \tablenotetext{(e)}{Number, percentage and its uncertainty of ``anti-Hale'' sunspots}
 \end{deluxetable}
\clearpage

\begin{deluxetable}{cclrrrr}
\tablewidth{0pt}
\tablecaption{Magnetic Flux [Mx]\label{tmb}}
\tablehead{
\colhead{Cycle\tablenotemark{(a)}} &
\colhead{Hem.\tablenotemark{(b)}} &
\colhead{Population\tablenotemark{(c)}} &
\colhead{$\sum\Phi\pm\Delta$\tablenotemark{(d)}}&
\colhead{$\bar\Phi\pm \Delta$\tablenotemark{(e)}} &
\colhead{$\phi^\circ$\tablenotemark{(f)}}
\\
\colhead{} &
\colhead{} &
\colhead{} &
\colhead{$\times10^{24}$ [Mx]} &
\colhead{$\times10^{21}$ [Mx]} &
\colhead{at $\Phi$ Peak}
}
\startdata
23 & N & Hale  &$13.04 \pm0.39$& $11.70\pm 0.35$& $14.7\pm1.1$ \\
        & & anti-Hale & $0.71\pm 0.07$ & $7.29\pm 0.74$ & $12.7\pm1.0$\\
 \cline{2-6}
 &S & Hale& $14.97\pm 0.41$& $11.31\pm 0.31$& $-15.5\pm1.0$\\
 & & anti-Hale & $1.18\pm0.10$& $8.80\pm 0.76$& $-13.6\pm1.1$\\
 \hline
 24 & N & Hale & $5.63\pm0.19$ & $6.57\pm0.22$& $13.8\pm1.5$\\
 & & anti-Hale  & $0.27\pm0.03$ & $4.45\pm0.57$ & $13.7\pm0.4$\\
 \cline{2-6}
 & S & Hale &$5.28\pm0.19$ &$ 7.17\pm 0.26$&$ -17.0\pm0.6$\\
 & & anti-Hale  & $0.25\pm0.03$&$ 3.96\pm0.50$& $-14.8\pm1.3$\\
\enddata
 \tablenotetext{(a)}{Solar cycle number}
 \tablenotetext{(b)}{Hemisphere}
 \tablenotetext{(c)}{Sunspot Populations: Hale and anti-Hale}
 \tablenotetext{(d)}{Cumulative sunspot magnetic flux and its uncertainty}
 \tablenotetext{(e)}{Average sunspot magnetic flux and its uncertainty}
 \tablenotetext{(f)}{Latitude at the peak of the cumulative magnetic flux and its uncertainty}
\end{deluxetable}
\clearpage

\begin{deluxetable}{rccrc}
\tablewidth{0pt}
\tablecaption{Sunspot Magnetic Flux and Pole Separation\label{tflux2s}}
\tablehead{
\colhead{Sunspot} &
\colhead{($\log_{10}\bar s\pm \Delta$)\tablenotemark{(a)}} &
\colhead{($k_s\pm\sigma$)\tablenotemark{(b)} } &
\colhead{($C_s\pm\sigma$)\tablenotemark{(c)}} &
\colhead{$r_{corr}$\tablenotemark{(d)} } \\
\colhead{Population} &
\colhead{} &
\colhead{} &
\colhead{} &
\colhead{} }
\startdata
Hale& $0.62\pm0.01$& $1.57\pm0.02$ & $20.82\pm 0.01$ & 0.75 \\
anti-Hale&$0.44\pm0.02$ &  $1.06\pm 0.09 $ & $ 21.17\pm 0.04$ & 0.54 \\
All & $0.53\pm0.01$ & $1.49\pm0.02$ & $20.88\pm0.02$ & 0.73
\enddata
\tablenotetext{(a)}{Logarithm of average pole separation [degree] and its uncertainty}
\tablenotetext{(b)}{Slope in the regression fitting parameter in Equation (\ref{eflux2s})}
\tablenotetext{(c)}{Intercept in the regression fitting parameter in Equation (\ref{eflux2s})}
\tablenotetext{(d)}{Pearson correlation coefficient determined between $\log\Phi$ and $\log s$}
\end{deluxetable}
\clearpage

\begin{deluxetable}{cclrrrrrrr}
\tablewidth{0pt}
\tablecaption{Sunspot Tilt Angle Summary\label{ttiltangles}}
\tablehead{
\colhead{Cycle\tablenotemark{(a)}} &
\colhead{Hem.\tablenotemark{(b)}} &
\colhead{Population\tablenotemark{(c)}} &
\multicolumn{3}{c}{Latitude\tablenotemark{(d)}} &
\colhead{} &
\multicolumn{3}{c}{Tilt Angle\tablenotemark{(e)}}\\
\cline{4-6} \cline{8-10} \\
\colhead{} &
\colhead{} &
\colhead{} &
\colhead{$\bar\phi$} &
\colhead{[$\phi$]} &
\colhead{$3\sigma_\phi$} &
\colhead{} & 
\colhead{$\bar\gamma$} &
\colhead{[$\gamma$]} &
\colhead{$3\sigma_\gamma$}
}
\startdata
23 & N &Hale &  15.63 & 15.06 &0.67 && 5.05 & 6.66 & 2.45 \\
    & & anti-Hale &14.58& 14.34 & 2.42&&-0.47 & 4.79& 16.00\\
    \cline{2-10}
    & S & Hale  & -15.83 & -15.17& 0.64 && 4.13 &5.61 & 2.30\\
    & & anti-Hale& -15.13 & -14.36 &2.04& &-7.23 & -9.13& 13.31\\
      \hline
24 & N & Hale & 14.56 & 14.11& 0.67&& 6.85 & 7.69 & 2.62 \\
     &  & anti-Hale & 13.84 & 13.30 &2.18& &-8.22&-8.46&21.76 \\
       \cline{2-10}
     & S & Hale & -16.37 & -16.43& 0.73&& 7.02 & 8.64 & 2.80 \\
      & & anti-Hale& -15.95 & -14.98 &2.72& &-9.01 & -3.37 &19.68
\enddata
 \tablenotetext{(a)}{Solar cycle number}
 \tablenotetext{(b)}{Hemisphere}
 \tablenotetext{(c)}{Sunspot Populations}
 \tablenotetext{(d)}{Sunspot latitude statistics: average ($\bar\phi$) and median ([$\phi$]) latitudes, and three times the standard deviation of mean ($\sigma_\phi$)}
 \tablenotetext{(e)}{Tilt angle statistics: average ($\bar\gamma$), median ([$\gamma$]) tilt angles, and three times the standard deviation of mean ($\sigma_\gamma$)}
 \end{deluxetable}
\clearpage

\begin{deluxetable}{crrcrrr}
\tablecaption{Joy's Law from all Hale Sunspots\label{tjoyslaw}}
\tablewidth{0pt}
\tablehead{
\colhead{Cycle\tablenotemark{(a)}} &
\multicolumn{2}{c}{23} &
\colhead{}&
\multicolumn{2}{c}{24} &
\colhead{Final} \\
\cline{2-3} \cline{5-6} \\
\colhead{Hemmisphere\tablenotemark{(b)}} &
\colhead{N} & \colhead{S} &
\colhead{} &
\colhead{N} & \colhead{S} &
\colhead{Joy's Law}
 }  
\startdata
 ($k_J\pm\sigma$)\tablenotemark{(c)}&  $0.36\pm 0.10$&$ -0.43\pm0.10$&& $0.38\pm0.12$ &$-0.36\pm0.13$ & $0.38\pm 0.05$\\
($C_J\pm\sigma$)\tablenotemark{(c')} & $-0.01\pm0.03$ & $-0.05\pm0.03$ &&$0.02\pm0.03$ &$ 0.01\pm0.04$ & $-0.01\pm0.02$\\
 ($k_\phi\pm\sigma$)\tablenotemark{(d)} & $0.36\pm0.11$&  $-0.45\pm0.10$ &&   $0.37\pm0.13 $& $ -0.39\pm0.14$& $0.39\pm 0.06$\\
 ($C_\phi\pm\sigma$)\tablenotemark{(d')} &  $-0.53\pm1.90$& $-3.00\pm1.73$ & &$1.43\pm2.14$& $0.56\pm2.49$ & $-0.66\pm1.00$\\
($k_\circ\pm\sigma$)\tablenotemark{(e)} & $21.46\pm6.62$ & $-27.32\pm5.97$& &$22.19\pm 8.05 $& $-23.87\pm8.49$ & $23.80\pm 3.51$ \\
 ($C_\circ\pm\sigma$)\tablenotemark{(e')}  & $-0.69\pm1.95$ & $-3.25\pm1.78$& &$1.31\pm2.19$& $0.34\pm2.55$& $-0.86\pm 1.03$
\enddata
 \tablenotetext{(a)}{Solar cycle number}
 \tablenotetext{(b)}{Hemisphere letters}
 \tablenotetext{(c)}{Fitted slope and its uncertainty for  Equation (\ref{joysin})}
 \tablenotetext{(c')}{Fitted constant and its uncertainty for Equation (\ref{joysin})}
 \tablenotetext{(d)}{Fitted slope and its uncertainty for Equation (\ref{joydirect})}
 \tablenotetext{(d')}{Fitted constant and its uncertainty for Equation (\ref{joydirect})}
 \tablenotetext{(e)}{Fitted slope and its uncertainty for  Equation (\ref{joystenflo})}
 \tablenotetext{(e')}{Fitted constant and its uncertainty for  Equation (\ref{joystenflo})}
 \end{deluxetable}
 
\begin{deluxetable}{cclccrrr}
\tablewidth{0pt}
\tablecaption{Hale Populations\label{haless}}
\tablehead{
\colhead{Cycle\tablenotemark{(a)}} &
\colhead{Hem.\tablenotemark{(b)}} &
\colhead{Population\tablenotemark{(c)}} &
\colhead{\#\tablenotemark{(d)}} &
\colhead{\%\tablenotemark{(e)}} &
\colhead{$\sum\Phi\pm\Delta$\tablenotemark{(f)}}&
\colhead{$\bar\Phi\pm \Delta$\tablenotemark{(g)}} &
\colhead{$\bar\phi^\circ$\tablenotemark{(h)}} 
\\
\colhead{} &
\colhead{} &
\colhead{} &
\colhead{} &
\colhead{} &
\colhead{$\times10^{24}$ [Mx]} &
\colhead{$\times10^{21}$ [Mx]} &
\colhead{}
}
\startdata
23 & N & Hale/normal  &700& $57.71\pm0.02$ &$8.8\pm 0.3$& $12.8\pm0.5$ & $16.2\pm0.6$\\
        & & Hale/inverted &415 & $34.21\pm0.02$&$4.1\pm 0.2$ & $9.8\pm 0.5$ & $14.7\pm0.7$\\
       \cline{2-8}
 &S & Hale/normal& 799&$54.84\pm0.02$& $9.3\pm 0.3$& $11.7\pm 0.4$& $-16.6\pm0.6$\\
 & & Hale/inverted & 524 & $35.96\pm0.02$&$5.7\pm0.3$& $10.8\pm 0.5$& $-14.7\pm0.6$ \\
 \hline
 24 & N & Hale/normal & 560 &$61.07\pm0.02$& $3.8\pm0.2$ & $6.8\pm0.3$&  $15.0\pm0.6$\\
 & & Hale/inverted  &297 &$32.39\pm0.01$& $1.8\pm0.1$ & $6.2\pm0.4$ & $13.8\pm0.8$\\
 \cline{2-8}
 & S & Hale/normal&495 &$62.03\pm0.02$&$3.7\pm0.2$ &$ 7.5\pm 0.3$&  $-16.9\pm0.8$\\
 & & Hale/inverted & 241  & $30.20\pm0.02$&$1.6\pm0.1$&$ 6.6\pm0.4$&  $-15.5\pm1.0$\\
\enddata
 \tablenotetext{(a)}{Solar cycle number}
 \tablenotetext{(b)}{Hemisphere}
 \tablenotetext{(c)}{Hale Sunspot sub-Population: Hale/normal and Hale/inverted}
 \tablenotetext{(d)}{Number of sunspots}
 \tablenotetext{(e)}{Percentage of sunspots with respective hemisphere and cycle and its uncertainty}
 \tablenotetext{(f)}{Cumulative sunspot magnetic flux and its uncertainty}
 \tablenotetext{(g)}{Average sunspot magnetic flux and its uncertainty}
 \tablenotetext{(h)}{Average latitude and its uncertainty}
 \end{deluxetable}

\clearpage

\clearpage

\clearpage

\end{document}